\documentclass[aps,prl,10pt,superscriptaddress,twocolumn]{revtex4-1}

\usepackage{tikz}
\usepackage[cmex10]{amsmath}
\usepackage{amssymb}
\usepackage{physics}
\usepackage{upgreek}
\usepackage{ulem}
\usepackage{titlesec}
\usepackage{siunitx}
\usepackage{here}
\usepackage{color}
\usepackage{pdfpages}
\usepackage{pgffor} 

\makeatletter
\AtBeginDocument{\let\LS@rot\@undefined}
\makeatother

\def\supplementfilename{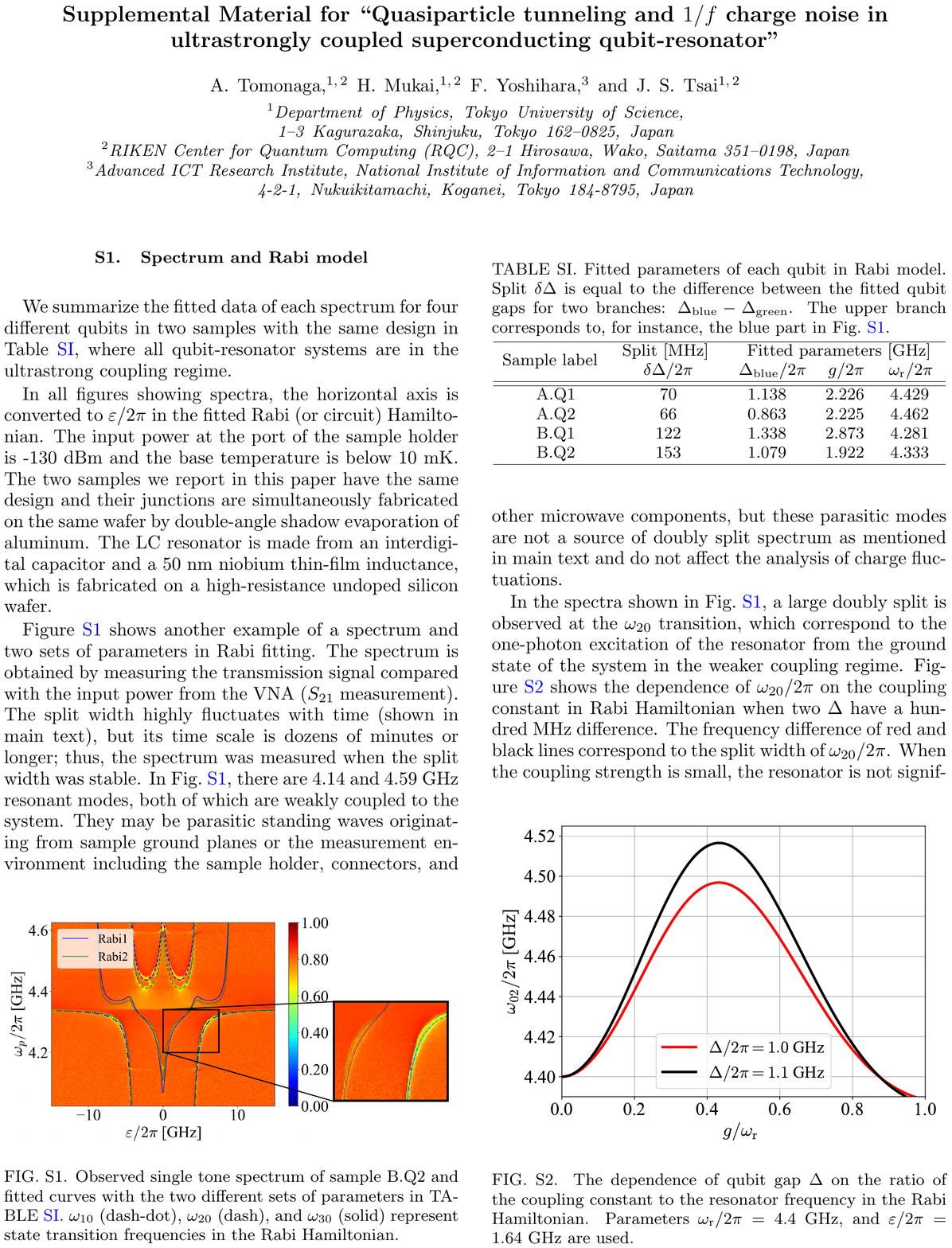}

\pdfximage{\supplementfilename}
\def\numbersupplementpages{\the\pdflastximagepages}

\newif\ifarXiv
\arXivtrue 

\newcommand{\vp}{\varphi}
\newcommand{\ej}{E_\mathrm{J}}
\newcommand{\ec}{E_\mathrm{c}}
\newcommand{\cj}{C_\mathrm{J}}
\newcommand{\ejec}{E_\mathrm{J}/E_\mathrm{c}}

\usepackage{hyperref}
\hypersetup{
    colorlinks=true,
    linktocpage=true,
    linkcolor=blue,
    filecolor=magenta,      
    urlcolor=cyan,
    bookmarks=true,}
    
\graphicspath{{images/}}

\newcommand{\addsec}[1]{\textcolor{black}{#1}}
\newcommand{\erasesec}[1]{}
\newcommand{\add}[1]{\textcolor{black}{#1}}
\newcommand{\erase}[1]{}

\begin{document}
\title{Quasiparticle tunneling and $1/f$ charge noise in ultrastrongly coupled superconducting qubit and resonator}

\author{A. Tomonaga}\email{akiyoshi.tomonaga@riken.jp}
\affiliation{Department of Physics, Tokyo University of Science, 1--3 Kagurazaka, Shinjuku, Tokyo 162--0825, Japan}
\affiliation{RIKEN Center for Quantum Computing (RQC), 2--1 Hirosawa, Wako, Saitama 351--0198, Japan}
\author{H. Mukai}
\affiliation{Department of Physics, Tokyo University of Science, 1--3 Kagurazaka, Shinjuku, Tokyo 162--0825, Japan}
\affiliation{RIKEN Center for Quantum Computing (RQC), 2--1 Hirosawa, Wako, Saitama 351--0198, Japan}
\author{F. Yoshihara}
\affiliation{Advanced ICT Research Institute, National Institute of Information and Communications Technology, 4-2-1, Nukuikitamachi, Koganei, Tokyo 184-8795, Japan}
\author{J. S. Tsai}\email{tsai@riken.jp}
\affiliation{Department of Physics, Tokyo University of Science, 1--3 Kagurazaka, Shinjuku, Tokyo 162--0825, Japan}
\affiliation{RIKEN Center for Quantum Computing (RQC), 2--1 Hirosawa, Wako, Saitama 351--0198, Japan}

\begin{abstract}
We report \add{on} an experimentally observed \erasesec{anomalous }doubly split spectrum and its split-width fluctuation \addsec{due to charge fluctuation} in an ultrastrongly coupled superconducting qubit and resonator. 
From an analysis of \add{the} circuit model Hamiltonian, we found that the doubly split spectrum and split-width fluctuation are caused by discrete charge hops due to quasiparticle tunneling\erase{s} and a continuous background charge fluctuation in islands of a flux qubit.
During 70 hours \add{of}\erase{in the} spectrum measurement, \add{the} split width fluctuate\add{d}\erase{s} but the middle frequency of the split \add{was}\erase{is} constant. 
This \add{observation}\erase{result} indicates that \add{the} quasiparticles in our device\erase{seem} mainly tunnel \add{in} one particular junction\add{, as expected from the energy difference between quasiparticle states, during this 70 hours}.
The background offset charge obtained from \add{the} split width has the $1/f$ noise characteristic.
\end{abstract}

\maketitle

\section{Introduction}
The study of various phenomena using superconducting artificial atoms and resonators with relatively freely selectable parameters compared with those of natural atoms has greatly contributed to our understanding of the physics in interacting light and matter as well as to the construction of quantum devices~\cite{blais_cavity_2004,kockum_ultrastrong_2019,kjaergaard_superconducting_2020,arute_quantum_2019}.
To control quantum states, microscopic noise behaviors and how they affect quantum devices should be understood. 
Thereby, several noises in superconducting circuits, such as charge, magnetic\add{-}flux, two-level-system (TLS), and quasiparticle noises, have been investigated in various cases~\cite{astafiev_quantum_2004,yoshihara_flux_2014,yan_spectroscopy_2012,muller_towards_2019,serniak_hot_2018,kwon_gate-based_2021}.

\erase{Charge noises originated from the poisoning of quasiparticles and the fluctuation in the environmental background electric field are one of the main noise sources in a superconducting qubit. These noises have mainly been observed in charge qubits~[9, 11-13].}
\add{A superconducting charge qubit~\cite{astafiev_quantum_2004,serniak_hot_2018} and a flux qubit~\cite{yoshihara_flux_2014,yan_spectroscopy_2012} are devices commonly used to evaluate the charge noise on a superconducting circuit, which is one of the main noise sources for the qubit coherence and originates from the poisoning of quasiparticles and the fluctuation in the environmental background electric field.
Quasiparticle poisoning sometimes induces a doubly split spectrum, which has mainly been observed and analyzed in charge qubits~\cite{christensen_anomalous_2019,riste_millisecond_2013,schreier_suppressing_2008,serniak_hot_2018}.}
Its effect on coherence has also been investigated~\cite{lutchyn_erratum_2007,catelani_relaxation_2011}.
\erase{The charging energy of a superconducting island separated by Josephson junctions (Cooper pair box: CPB) varies with the external gate voltage with a periodicity of $2e$~[16, 17]. The tunneling of existing non-equilibrium quasiparticles changes the excess charge on the island switching between two energy states (even- and odd-charge parity states) because of the $2e$ periodicity of the gate-induced charge. The fluctuation between these classical two energy states exhibits a doubly split spectrum. The energy difference between the two states is suppressed exponentially in terms of the ratio of the Josephson (current) energy ($E_\mathrm{J}=\hbar I_\mathrm{c}/2e$) to the charge energy ($E_\mathrm{c}=e^2/2\cj$) of the Josephson junction, where $I_\mathrm{c}$ and $\cj$ represent the critical current and capacitance of the Josephson junction, respectively~[13, 18].}

\add{The qubit energy structure is characterized by the circuit parameters especially the Josephson (current) energy ($E_\mathrm{J}=\hbar I_\mathrm{c}/2e$) and charge energy ($E_\mathrm{c}=e^2/2\cj$), where $I_\mathrm{c}$ and $\cj$ represent the critical current and capacitance of the Josephson junction, respectively~\cite{nakamura_coherent_1999,koch_charge-insensitive_2007}.
A charge qubit [transmon and Cooper pair box (CPB)] composed of a superconducting island separated by a Josephson junction has Hamiltonian $\mathcal{H}_\mathrm{cq}=4E_\mathrm{c}(n-n_\mathrm{g})^2-E_\mathrm{J}\cos{\theta}$, where $n$ is the number of excess Cooper pairs on the island, $n_\mathrm{g}$ is the offset charge induced by the gate charge or electrostatic environment, and $\theta$ is the superconducting phase difference across the junction.
Thus, the eigenenergies of a charge qubit vary with the external gate voltage with a periodicity of $2e$~\cite{nakamura_coherent_1999,yamamoto_parity_2006}.
The tunneling of existing non-equilibrium quasiparticles changes the excess charge on the island $n_\mathrm{g}$, switching it between two energy states (even- and odd-charge parity states), \addsec{i.e. a single quasiparticle changes $n_\mathrm{g}$ with $1e$.}
The fluctuation between these two energy states exhibits a doubly split spectrum. 
The energy difference between the two states is suppressed exponentially with increasing ratio $\ejec$~\cite{schreier_suppressing_2008,koch_charge-insensitive_2007}.}

\begin{figure*}
\centering
\includegraphics[width=180mm]{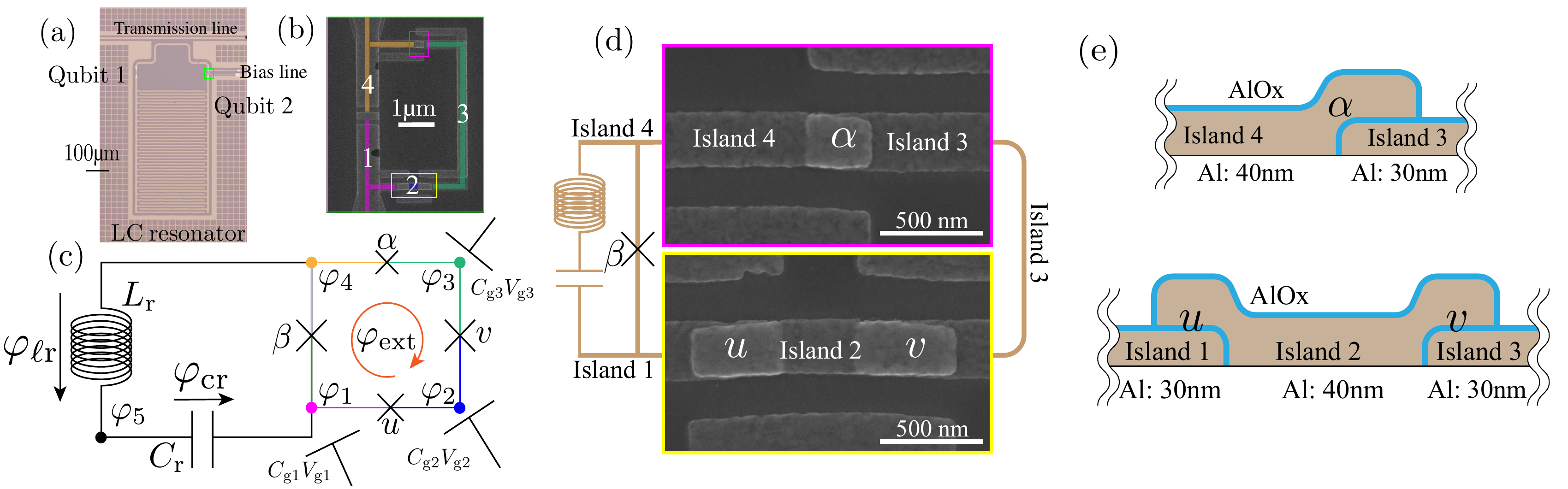}
\caption{(a) Optical microscope image of measured sample A. The sample holder has a coil to bias a uniform magnetic field from the back face of the chip. 
\add{Qubit 2 has a local bias line to change the magnetic flux for the qubit loop.}
(b) False-color SEM image of qubit 2 in (a).\erase{The size of island 3 is $2.6~\si{\micro}\si{\meter}^2$ and that of island 2 is $0.1~\si{\micro}\si{\meter}^2$.}
(c) \add{Circuit diagram of sample A, in which the colors of the qubit loop correspond to the colors of the sample shown in (b), with gate capacitors used to calculate the effect of the gate-induced island charge $C_{\mathrm{g}i}V_{\mathrm{g}i}$ ($i=1,2,3$).}\erase{Circuit schematic image colored mach to (b) with gate capacitors $C_{\mathrm{g}i}V_{\mathrm{g}i}$ ($i=1,2,3$) to calculate the effect of the island charge.} 
In a real system, these islands are subjected to a background charge from \add{environment} such as the ground pad, dielectric wafer, and islands themselves\add{, which} can also store electrons.
\add{(d) Enlarged SEM image of $\alpha$-, $u$-, and $v$-junction areas in (a). The size of the $v$-junction obtained from the SEM image is 0.12 $\si{\micro\meter}^2$ and the area ratios of the junctions to the $v$-junction are $\alpha=0.66$, $\beta=1.98$, and $u=0.91$. 
The size of island 2 is $0.32~\si{\micro\meter}^2$ and that of island 3 is $2.81~\si{\micro\meter}^2$.
(e) Schematic cross sections of the $\alpha$-, $u$-, and $v$-junctions.}
}
\label{Fig1}
\end{figure*}

\add{A flux qubit, which is composed of a superconducting loop including several Josephson junctions, is usually controlled by magnetic flux, but the eigenenergies of the flux qubit also depend on the gate charge in an island isolated by two Josephson junctions.}
The charge dependence in the flux qubit can be understood as the Aharonov--Casher effect, known as the dual of the Aharonov--Bohm effect, caused by\erase{a} quantum interference between moving magnetic dipoles affected by the electric field~\cite{aharonov_topological_1984,friedman_aharonov-casher-effect_2002,de_graaf_charge_2018}.
\add{The}\erase{A} phase $\varphi$ of the wave function of a moving magnetic dipole $\boldsymbol{\upmu}$ on path $\mathcal{P}_{\vb{x}}$ with electric field $\mathcal{E}$ is described by $\varphi=(1/\hbar c^2)\int_\mathcal{P} \qty(\vb{\mathcal{E}} \times \boldsymbol{\upmu})\cdot\dd{\vb{x}}$.  
\add{That is}\erase{In other words}, the flux qubit is subject to electric field fluctuations via the Aharonov--Casher effect, \add{where the interference of magnetic flux trajectories occurs around the charge on superconducting islands of the flux qubit, and the tunneling rate of the magnetic flux changes}\erase{in which a magnetic flux in the qubit loop interferes with itself}, resulting in fluctuation of\erase{the} eigenenergies.
A doubly split spectrum due to this effect in flux qubits has been reported~\cite{bertet_dephasing_2005,stern_flux_2014,bal_dynamics_2015}.
In Ref.~\onlinecite{bal_dynamics_2015}, the transition rate between two energy states was obtained by taking the correlation of relaxations from each state, and it was concluded that the results were consistent with the theoretically predicted value from the effect of non-equilibrium quasiparticles.

In this \add{paper}\erase{letter}, we report on the observed \erasesec{anomalous }doubly split spectrum and its fluctuations in a superconducting flux qubit ultrastrongly coupled with a resonator.
No such\erase{kind of} doubly split spectrum has\erase{not} been observed in an ultrastrongly coupled system.
The ultrastrong $(0.1\lesssim \{g/\omega_\mathrm{r}$, $g/\add{\omega_\mathrm{q}}\erase{\Delta}\}< 1)$ and deep-strong $(1\lesssim \{g/\omega_\mathrm{r}$, $g/\add{\omega_\mathrm{q}}\erase{\Delta}\})$ coupling regimes have recently been implemented in superconducting circuits ($\hbar g$, coupling energy; $\add{\hbar}\omega_\mathrm{r}\erase{/2\pi}$, resonator \add{energy}\erase{frequency}; and $\add{\hbar\omega_\mathrm{q}}\erase{\Delta}$, qubit \add{energy}\erase{gap})~\cite{kockum_ultrastrong_2019,forn-diaz_ultrastrong_2019}.
These systems are expected to play an important role as tools for quantum information processing such as \add{in} ultrafast two-qubit\erase{s} phase gates~\cite{romero_ultrafast_2012}, quantum computation~\cite{stassi_scalable_2020,wang_ultrafast_2017,nataf_protected_2011,kyaw_creation_2015}\erase{quantum simulation~\cite{braumuller_analog_2017}}, quantum annealing~\cite{mukai_superconducting_2019,pino_quantum_2018}, and quantum memory~\cite{stassi_long-lasting_2018}.

\addsec{In section \ref{sec:circuitHamiltonian}, to investigate the charge effect on an ultrastrongly coupled system and understand the microscopic charge noise behavior, we implement a circuit Hamiltonian.}
\erasesec{We implement the circuit Hamiltonian, which enable\add{s us} to investigate the charge effect on an ultrastrongly coupled system, to understand the microscopic charge noise behavior.}
By examining the Rabi and circuit model Hamiltonians and the dependence of their spectrum on the island charge, \add{we found that} the doubly split spectrum in our devices was also attributed to quasiparticle tunneling.
Moreover, we also found an effect of the environmental electric field noise on the energy level of a flux qubit with a longer time scale \add{in addition to}\erase{, besides} the presence of quasiparticle fluctuation.
Although \add{the} observed charge noises could not be enhanced or directly related to the nature of ultrastrong coupling, the shape of the spectrum and the highly entangled states of the ultrastrong coupling helped \add{us}\erase{to} evaluate these noises~\cite{noauthor_see_nodate}.

\begin{figure*}[t!]
\centering
\includegraphics[width=180mm]{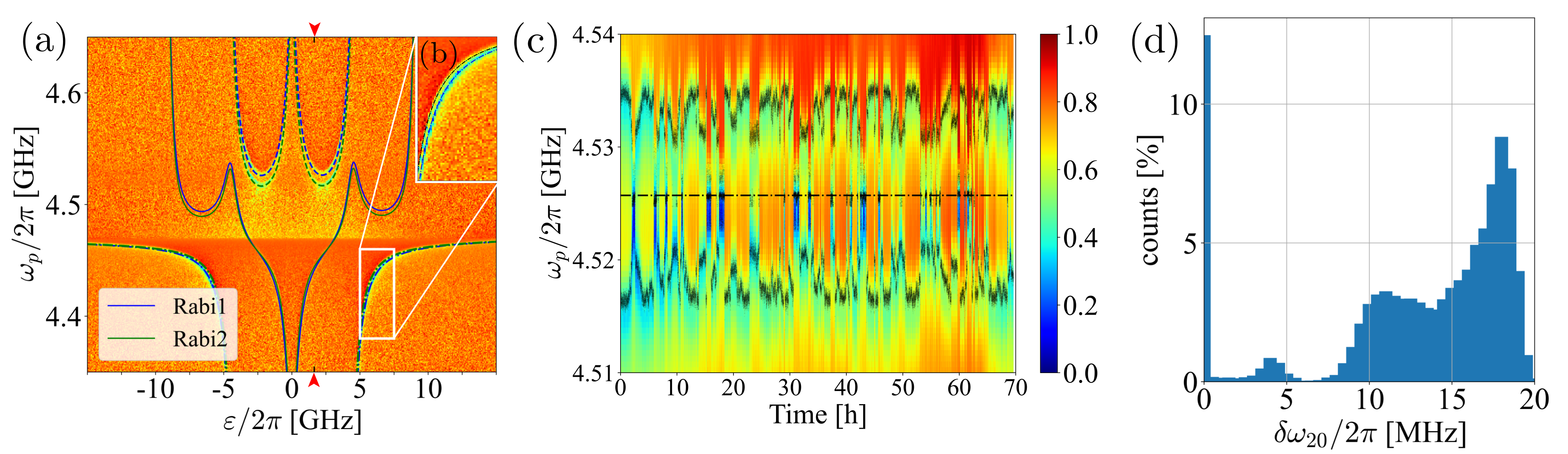}
\caption{
(a) Observed single\add{-}tone spectrum of sample A.Q2 and curves fitted with $\Delta_\mathrm{blue}/2\pi=\add{0.863}\erase{0.847}$, $\Delta_\mathrm{green}/2\pi=0.797$, $g/2\pi=2.225$, and $\omega_\mathrm{r}/2\pi=4.462~\si{\giga\hertz}$.
$\omega_{10}$ (dash-dot), $\omega_{20}$ (dash), and $\omega_{31}$ (solid) represent state transition frequencies in the Rabi Hamiltonian. The transition frequencies of two Rabi Hamiltonians with different sets of parameters are plotted in blue and green.
(b) Enlarged image of (a) \add{in}\erase{at} the area $\varepsilon/2\pi\add{~[\si{\giga\hertz}]}: [5.0,\,7.5]$ and $\omega_p/2\pi~[\si{\giga\hertz}]: [4.38,\,4.46]$.
(c) Result of 70-hours repeated measurements with a single tone at $\varepsilon/2\pi=1.64~\si{\giga\hertz}$ [red arrows in (a)] with three-second intervals. 
Small black dots are the obtained peak positions of the upper and lower branches of each measurement trace. The dash--dot line represents the averaged middle frequency of the split\add{, which is mostly at $\bar{f}_\mathrm{m}=4.526$ GHz}.
The time for which microwaves are applied at each frequency is more than \add{$20~\si{\milli\second}$}\erase{0.02~s}, which is the inverse of the $50~\si{\hertz}$ IF bandwidth of the vector network analyzer.
(d) Histogram of split width $\delta\omega_{20}\add{/2\pi}$ obtained from (c) with 500$~\si{\kilo\hertz}$ binning~\cite{noauthor_see_nodate}. 
}
\label{Fig2}
\end{figure*}

\section{\add{Spectrum splitting in Rabi model}}\label{sec:Spectrum splitting in Rabi model}

The Rabi model is a phenomenological model \add{used} to describe the interacting spin and electromagnetic field. It\erase{and} has been widely adopted as the representation of a quantum system in which a qubit and a resonator interact including ultrastrong and deep\add{-}strong coupling regimes~\cite{yoshihara_superconducting_2017,yoshihara_characteristic_2017,niemczyk_circuit_2010}.
The Hamiltonian of the Rabi model is given by
\begin{align}
    \mathcal{H}_\mathrm{\add{Rabi}\erase{rabi}}/\hbar \! = \!
    \frac{1}{2}\qty(\varepsilon\sigma_\mathrm{z} \! + \! \Delta\sigma_\mathrm{x}) \! + \! \omega_\mathrm{r} \qty(\!a^\dagger a \! + \!\frac{1}{2}\!) \!
    \! + \! g\sigma_\mathrm{z}\qty(a^\dagger \!+ a) \,,
\label{eq:Rabi}
\end{align}
where \add{$\hbar\Delta$ and} $\hbar\varepsilon$ represent \add{the qubit energy gap and} the energy of \add{the}\erase{a} loop current of a flux qubit corresponding to the external field for a spin, respectively. 
\add{The qubit energy $\hbar\omega_\mathrm{q}$ can be written as $\hbar\omega_\mathrm{q}=\hbar\sqrt{\varepsilon^2+\Delta^2}$.}
In the measured sample shown in Fig.~\ref{Fig1}(a), a flux qubit is connected to a lumped element (LC) resonator via a Josephson junction that separates a shared line. Considering the junction (called the $\beta$-junction) as the coupling inductance between the qubit and the resonator, the Hamiltonian of the coupling is derived as $L_\beta I_\mathrm{zpf}I_\mathrm{q}\sigma_z(a^\dagger+a)$ from an analogy of the classical circuit, where $L_\beta$, $I_\mathrm{zpf}$, and $I_\mathrm{q}$ are the inductance of the $\beta$-junction, the zero-point fluctuation current of the LC resonator, and the screening current of the qubit, respectively.


We measure four qubits in two samples with the same design in the ultrastrong coupling regime\add{, as}\erase{which} summarized in the Supplemental Materials~\cite{noauthor_see_nodate}.
The spectrum obtained from one of the sample\add{s} (labeled A.Q2) is shown in Fig.\add{~\ref{Fig2}(a)}\erase{\ref{Fig1}(d)}, where the energy absorption lines form an \erasesec{anomalous }doubly split shape.
We first fit three energy absorption lines (blue) to \add{state transition frequencies} $\omega_{ij}$ \add{($i,j\in\qty{0,1,2,3}$)}\erase{in Eq.~\eqref{eq:Rabi}} including the upper branch of $\omega_{20}$ (blue dash line), then fit the other split branch of $\omega_{20}$ (green dash line) with $\Delta$ as a \add{fitting parameter}\erase{variable}; the other parameters are the same as the blue lines, where $\add{\hbar}\omega_{ij}$ \add{corresponds to the energy difference}\erase{represents the state transition frequency} between the $i$th and $j$th eigenstates of the Rabi model \add{Hamiltonian Eq.~\eqref{eq:Rabi}}.
The fitting with two sets of parameters well reproduced the experimental results including the small splits in $\omega_{10}$ \add{shown in Fig.~\ref{Fig2}(b)}. Any parameter sets of a single Rabi Hamiltonian cannot represent the measured spectrum.
Note that in the measured circuit, ultrastrong coupling enables\erase{to see} the $\Delta$ ($\omega_{10}$) splitting in $\omega_{20}$ \add{to be observed by}\erase{with} single-tone spectroscopy.
In \add{the} case of weaker coupling, two\add{-}tone spectroscopy is \add{required}\erase{needed} to \add{observe}\erase{see} such a low \add{frequency}\erase{energy} of $\Delta\add{/2\pi}$~\cite{noauthor_see_nodate}.
\add{Moreover}\erase{Also}, to obtain the transition frequencies from the spectrum, especially the split $\omega_{20}\add{/2\pi}$ (dash lines) peaks, we used image processing to extract ridge structures from the noisy three-dimensional image data \add{owing}\erase{due} to their close frequencies and the large noise floor~\cite{walt_scikit-image_2014,noauthor_see_nodate}.
In the fitting function in Eq.~\eqref{eq:Rabi}, $\omega_\mathrm{r}$ depends on the qubit state
and $\varepsilon$ via the $\beta$-junction~\cite{yoshihara_superconducting_2017}.

The fact that two different sets of parameters reproduce the experimental result using Eq.~\eqref{eq:Rabi} suggests the existence of a perturbation that splits the energy level $\add{\hbar}\Delta$ of the qubit and/or the existence of a noise that classically fluctuates the parameter. 
In the former case, where the qubit--resonator system couples to a TLS and/or a parasitic (boson) mode on the sample, \add{this}\erase{a} simple coupled model of the Hamiltonian does not reproduce all the splits in Fig.~\add{\ref{Fig2}(a)}\erase{1(d)} \add{including the fluctuating TLS and/or bosonic mode~\cite{schlor_correlating_2019}}.
The dressed states in Eq.~\eqref{eq:Rabi} also do not give rise to the doubly split shape of the measured spectrum. 
In addition, although this system has two qubits coupled to a common resonator, we can deal with the other qubit as a classical inductance when it is biased far from \add{its}\erase{the} optimal point~\cite{noauthor_see_nodate}.
To confirm the possibility of the latter case, we measure the split $\omega_{20}$ 84,000 times at the same \add{fixed} bias point in Fig.~\add{\ref{Fig2}(a)}\erase{1(d)}, and the result is shown in Fig.~\ref{Fig2}(c)\erase{1(f)}. 
The split of $\omega_{20}$ varies with time \add{and appears to have no}\erase{without an} obvious periodic structure\erase{.}\add{, which indicates a existence of fluctuator that changes the qubit parameter $\Delta$}.\erase{Thereby, we conclude that the splitting of the energy spectrum is caused by a noise, which induces the fluctuation in the qubit parameter $\Delta$. }

Consequently, the doubly split shape of the measured spectrum can only be observed if the system moves back and forth between two states with a sufficiently \add{shorter}\erase{faster} time constant than the measurement time of the vector network analyzer (VNA) that we used.
Possible factors that can change the energy of the system are the magnetic flux through \add{the} loop and the charge on\erase{an} island\add{s}. 
In the case of a magnetic flux noise, the spectrum will fluctuate in the x-axis ($\varepsilon$) direction; thus, the middle frequency of the split should fluctuate. 
However, the middle frequency in Fig.~\add{\ref{Fig2}(c)}\erase{1(f)} is almost constant and only the width of the split fluctuates. 
Thus, this fluctuation should not originate from a magnetic flux noise. 
\erase{In addition, a system coupled to a fluctuating TLS could fluctuate $\Delta$ by a dispersive shift. However, as shown in Rabi fits Fig~\ref{Fig1}(d), no such large shift of $50~\si{\mega\hertz}$ is expected to occur by the dispersive shift~\cite{schlor_correlating_2019}.}

\begin{figure*}[t!]
\centering
\includegraphics[width=180mm]{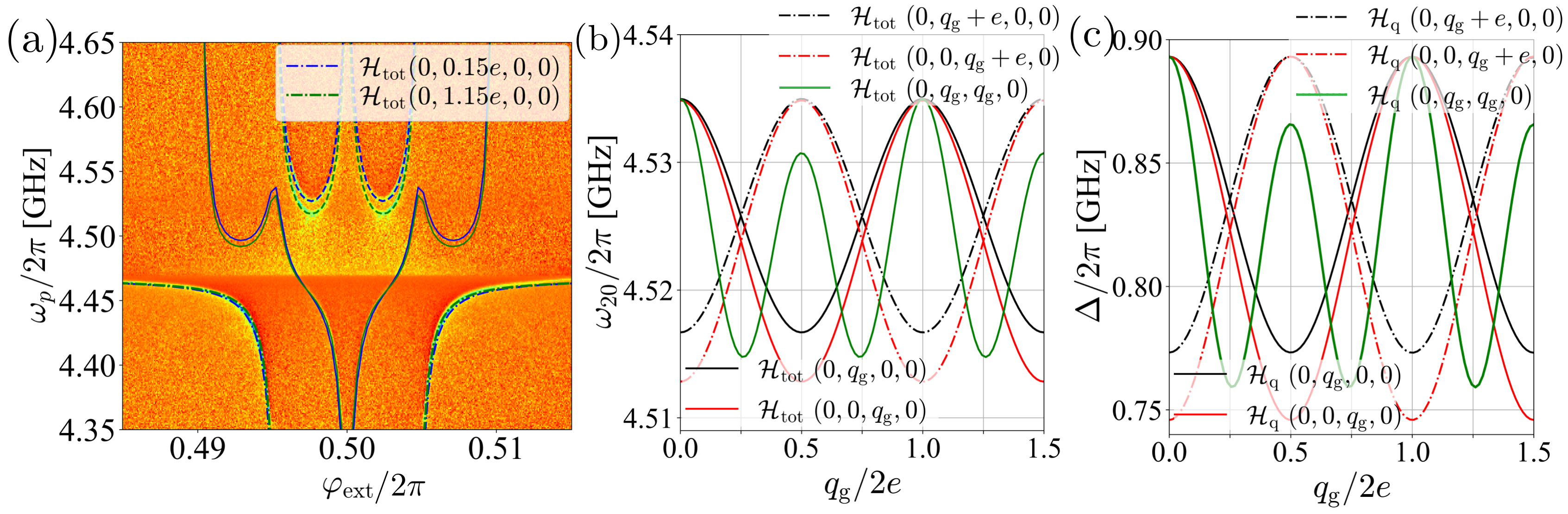}
\caption{
(a) Spectrum of sample A.Q2 \add{and fitted curves using the two charge states, $\mathcal{H}_\mathrm{tot}(0,0.15e,0,0)$ (blue) and $\mathcal{H}_\mathrm{tot}(0,1.15e,0,0)$ (green), with parameters $\ej/h=124~\si{\giga\hertz}$, $\ec/h=4.02~\si{\giga\hertz}$ ($\ejec=30.9$), $\omega_\mathrm{r}/2\pi=4.68~\si{\giga\hertz}$, $L_\mathrm{r}=6.84~\si{\nano\henry}$, $\alpha=0.76$, $\beta=2.02$, $u=0.90$, and $\eta_{1,2,3}=0.12$.}\erase{fitted by the circuit Hamiltonian using the two parity states $\mathcal{H}_\mathrm{tot}(0,0,0,0)$ (blue) and $\mathcal{H}_\mathrm{tot}(0,0,e,0)$ (green) with parameters $\ej/h=120~\si{\giga\hertz}$, $\ec/h=3.91~\si{\giga\hertz}$ ($\ejec=30.7$), $\omega_\mathrm{r}/2\pi=4.75~\si{\giga\hertz}$, $L_\mathrm{r}=3.34~\si{\nano\henry}$, $\alpha=0.83$, $\beta=3.10$, and $u=0.99$.}
$\omega_{10}$ (dash--dot), $\omega_{20}$ (dash), and $\omega_{31}$ (solid) represent the state transition frequencies in the circuit Hamiltonian.
\add{This spectrum is obtained by sweeping the current of the on-chip bias line for 3.5 hours.}
(b)(c)\add{Dependence of $\omega_{20}/2\pi$ at $\vp_\mathrm{ext}/2\pi=0.5018$ and qubit energy gap frequency $\Delta/2\pi$ on charge offset of islands 2 and 3 from the numerical calculation of Eqs.~\eqref{eq:TotalHami} and \eqref{eq:Hq} with the same parameters as those in (a), respectively.}\erase{Dependence of qubit gap $\Delta$ at $\vp_\mathrm{ext}=0.5$ and $\omega_{20}/2\pi$ at $\vp_\mathrm{ext}=0.5014$ on charge offset of islands 2 and 3 from numerical calculation of Eq.~\eqref{eq:Hq} with the parameters in (a).}
Dash-dot lines represent the case that an excess number of quasiparticles exist in island 2 or 3.
}
\label{Fig3}
\end{figure*}

\section{\add{Charging effect on flux qubit}}\label{sec:circuitHamiltonian}

\erase{A charge noise can fluctuate the spectrum in y-axis direction changing the parameter of the qubit.}
To determine which circuit parameters change $\Delta$ in the Rabi Hamiltonian and how the system produces the spectrum in \add{Figs.~\ref{Fig2}(a)--(c)}\erase{Figs.1(d) and (f)}, we solve a circuit Hamiltonian ~\cite{noauthor_see_nodate,billangeon_circuit-qed-based_2015,robertson_quantum_2006,peropadre_nonequilibrium_2013,bourassa_ultrastrong_2009}.
\add{On the basis of}\erase{In} Fig.~\ref{Fig1}(c), to calculate the charge dependence of each island separated by Josephson junctions, we use the node fluxes $\phi_i\equiv\varphi_i\times\Phi_0/2\pi$ of the islands as the calculation basis and define $\varphi_4\equiv0$ as the origin of the calculation basis\add{, where $\Phi_0$ is the flux quantum}.

From these definitions, using $I_\mathrm{zpf}=\sqrt{\hbar\omega_\mathrm{r}/2L_\mathrm{r}}$ and $\omega_\mathrm{r}=1/\sqrt{L_\mathrm{r}C_\mathrm{r}}$, \add{we derive} the Hamiltonian of the resonator $\mathcal{H}_\mathrm{r}$\erase{is derived} from Kirchhoff’s
voltage law for a closed loop containing $L_\mathrm{r}$ and $C_\mathrm{r}$:
\begin{align}
    \mathcal{H}_\mathrm{r}&=\hbar\omega_\mathrm{r}\qty(a^\dagger a+\frac{1}{2}) - I_\mathrm{zpf}\phi_1\qty(a^\dagger+a)\erase{+\frac{1}{2L_\mathrm{r}}\phi_1^2} \,,
    \label{eq:Hr}
\end{align}
where the annihilation and creation operators are $a\equiv(\phi_\mathrm{cr}-iZ_\mathrm{r} q_\mathrm{cr})/\sqrt{2\hbar Z_\mathrm{r}}$ and $a^\dagger\equiv(\phi_\mathrm{cr}+iZ_\mathrm{r} q_\mathrm{cr})/\sqrt{2\hbar Z_\mathrm{r}}$, respectively, with the characteristic impedance $Z_\mathrm{r}=\sqrt{L_\mathrm{r}/C_\mathrm{r}}$ \add{and $q_\mathrm{cr}$ as the canonical conjugate for $\dot{\phi}_\mathrm{cr}$}.
When the flux across the capacitance $C_\mathrm{r}$ ($\varphi_\mathrm{cr}$) is assumed to be the basis of the resonator, the second term of Eq.~\eqref{eq:Hr} is the coupling term~\cite{yoshihara_hamiltonian_2020,noauthor_see_nodate}.

We also define the Josephson energy of the $v$-junction as $E_\mathrm{J}$; $\alpha, \beta$, and $u$ as the ratios of junction areas to the $v$-junction area; \add{and} $\eta_i\equiv C_{\mathrm{g}i}/\cj$ as the ratio of the gate capacitance to the $v$-junction capacitance. 
The gate charge vector in Fig.~\ref{Fig1}(c), $2e\tilde{\vb{q}}_\mathrm{g}  \equiv (\, q_\mathrm{g1} \,\,\, q_\mathrm{g2} \,\,\, q_\mathrm{g3} \,)^\mathrm{T}$, \add{is}\erase{are} taken into account as the offset values of the charge basis $2e\tilde{\vb{q}}'=2e(\tilde{\vb{q}}+\tilde{\vb{q}}_\mathrm{g})$, which represent the sum of \add{the} island charges, where $2e\tilde{q}_{\mathrm{g}i}=C_i V_{\mathrm{g}i}$, $2e\tilde{\vb{q}}\equiv (\,q_1 \,\,\, q_2 \,\,\, q_3\,)^\mathrm{T}$, and $q_i$ is the canonical conjugate for $\phi_i$.
\add{Thereby}\erase{Thus}, we obtain the total Hamiltonian of the circuit as
\begin{align}
    \mathcal{H}_{\mathrm{tot}}\add{(\vb{q}_\mathrm{g},\varphi_\mathrm{ext})}=4E_\mathrm{c}\tilde{\vb{q}}'{}^\mathrm{T} \tilde{\vb{M}}^{-1}\tilde{\vb{q}}'
    \add{+E_\mathrm{Lr}\varphi_1^2}
    +\mathcal{U}_\mathrm{J}+\mathcal{H}_\mathrm{r} \,,
    \label{eq:TotalHami}
\end{align}
where $\tilde{\vb{M}}$ is the normalized mass matrix~\cite{noauthor_see_nodate}\add{, $E_\mathrm{Lr}=\Phi_0^2/(2L_\mathrm{r})$,} and $\mathcal{U}_\mathrm{J}$ is the qubit potential energy\add{, which is described by
\begin{align}
        \!\!\! \mathcal{U}_\mathrm{J}(\varphi_\mathrm{ext})=&-\ej
        \left[\,
        \beta\cos{(\varphi_1)}
        +u\cos{(\varphi_2-\varphi_1)} 
        \right. \notag \\ 
        &\,\,\quad\quad\left.
        +\cos{(\varphi_3-\varphi_2)}
        +\alpha\cos{(\varphi_\mathrm{ext}-\varphi_3)}
        \right] \,.
    \label{eq:uj}
\end{align}
}The numerical diagonalization of the total Hamiltonian in Eq.~\eqref{eq:TotalHami} gives the eigenenergies and eigenvectors of the circuit.
\add{We define $\epsilon_i (q_\mathrm{g1},q_\mathrm{g2},q_\mathrm{g3},q_\mathrm{g4})$ as the $i$th eigenenergy of $\mathcal{H}_{\mathrm{tot}}(q_\mathrm{g1},q_\mathrm{g2},q_\mathrm{g3},q_\mathrm{g4})$.}
\add{The state transition energy $\hbar\omega_{ij}$ is expressed by the difference \add{between the}\erase{of} $i$th and $j$th state eigenenergies $\hbar\omega_{ij}=\epsilon_j-\epsilon_i$.
Here, \add{the} transition between different charge states is not considered.}
\erase{Figure.2(a) shows a fitting result obtained using the circuit Hamiltonian with two parity states $\mathcal{H}_{\mathrm{tot}}(0,0,0,0)$ and $\mathcal{H}_{\mathrm{tot}}(0,0,e,0)$ in $\mathcal{H}_{\mathrm{tot}}(q_\mathrm{g1},q_\mathrm{g2},q_\mathrm{g3},q_\mathrm{g4})$.}

Figure~\add{\ref{Fig3}(a) shows a fitting result obtained using the circuit Hamiltonian with two charge states, $\mathcal{H}_{\mathrm{tot}}(0,0.15e,0,0)$ and $\mathcal{H}_{\mathrm{tot}}(0,1.15e,0,0)$, in $\mathcal{H}_{\mathrm{tot}}(q_\mathrm{g1},q_\mathrm{g2},q_\mathrm{g3},q_\mathrm{g4})$ with the finite offset gate charge 0.15$e$ for island 2.
As we discuss in sections~\ref{sec:QP} and \ref{sec:background}, quasiparticle poisoning and background charge noise mainly affect island 2. Thus, we use $q_\mathrm{g2}$ as one of the fitting parameters with two charge parities, $(0,q_\mathrm{g2},0,0)$ and $(0,q_\mathrm{g2}+e,0,0)$. 
The other fitted parameters are $\ej/h=124~\si{\giga\hertz}$, $\ec/h=4.02~\si{\giga\hertz}$ ($\ejec=30.9$), $\omega_\mathrm{r}/2\pi=4.68~\si{\giga\hertz}$, $L_\mathrm{r}=6.84~\si{\nano\henry}$, $\alpha=0.76$, $\beta=2.02$, $u=0.90$, and $\eta_{1,2,3}=0.12$.
Stray capacitances $\eta_i$ [gate capacitance in Fig.~\ref{Fig1}(c)] are also considered in this fitting. The value of $\alpha=0.76$ obtained from fitting is larger than that ($0.66$) obtained from the SEM image [Fig.~\ref{Fig1}(d)]. 
The larger value of $\alpha$ from the fitting can be explained by the effect of stray capacitances and loop inductance.
Also, the dependence of the qubit state transition frequency $\omega_{20}/2\pi$ on the gate charge at $\varphi_\mathrm{ext}/2\pi=0.5018$ obtained using the total Hamiltonian Eq.~\eqref{eq:TotalHami} is shown in Fig.~\ref{Fig3}(b).
In the fitting in Fig.~\ref{Fig3}(a), we use two constraints to reproduce not only Fig.~\ref{Fig3}(a) but also Fig.~\ref{Fig2}(c).
First, the frequency $\omega_{20}\add{/2\pi}$ in $\mathcal{H}_\mathrm{tot}(0,0.5e,0,0)$ at $\varphi_\mathrm{ext}\add{/2\pi}=0.5018$ corresponds to the averaged middle frequency of splitting $\bar{f}_\mathrm{m}=4.526$ GHz in Fig.~\ref{Fig2}(c).
Second, the maximum split-width around 18 MHz in Figs.~\ref{Fig2}(c) and (d) corresponds to the difference in $\omega_{20}/2\pi$ between the two charge parity states $\mathcal{H}_\mathrm{tot}(0,0,0,0)$ and $\mathcal{H}_\mathrm{tot}(0,e,0,0)$ at $\varphi_\mathrm{ext}/\add{2\pi}=0.5018$.}

\add{The qubit energy gap $\hbar\Delta(\vb{q}_\mathrm{g})$ in the circuit model is defined as the difference between the two lowest eigenenergies of the qubit Hamiltonian at $\varphi_\mathrm{ext}\add{/2\pi}=0.5$:}
\erase{The dependence of the qubit energy gap on the gate charge shown in Fig.~\ref{Fig2}(b) is obtained from the numerical diagonalization of the qubit Hamiltonian $\mathcal{H}_{\mathrm{q}}(q_\mathrm{g1},q_\mathrm{g2},q_\mathrm{g3},q_\mathrm{g4})$, where}
\begin{align}
    \mathcal{H}_{\mathrm{q}}\add{(\vb{q}_\mathrm{g},\varphi_\mathrm{ext})} \equiv 4E_\mathrm{c}\tilde{\vb{q}}'{}^\mathrm{T} \tilde{\vb{M}}^{-1}\tilde{\vb{q}}'
   \erase{+\frac{\phi_1^2}{2L_\mathrm{r}}}
    \add{+E_\mathrm{Lr}\varphi_1^2}
    +\mathcal{U}_\mathrm{J} \,.
    \label{eq:Hq}
\end{align}
\add{The dependence of $\Delta$ on the gate charge shown in Fig.~\ref{Fig3}(c) is obtained from the numerical diagonalization of $\mathcal{H}_\mathrm{q}(q_\mathrm{g1},q_\mathrm{g2},q_\mathrm{g3},q_\mathrm{g4})$ with the same parameters as those in Fig.~\ref{Fig3}(a).}
\begin{figure*}[t!]
\centering
\includegraphics[width=180mm]{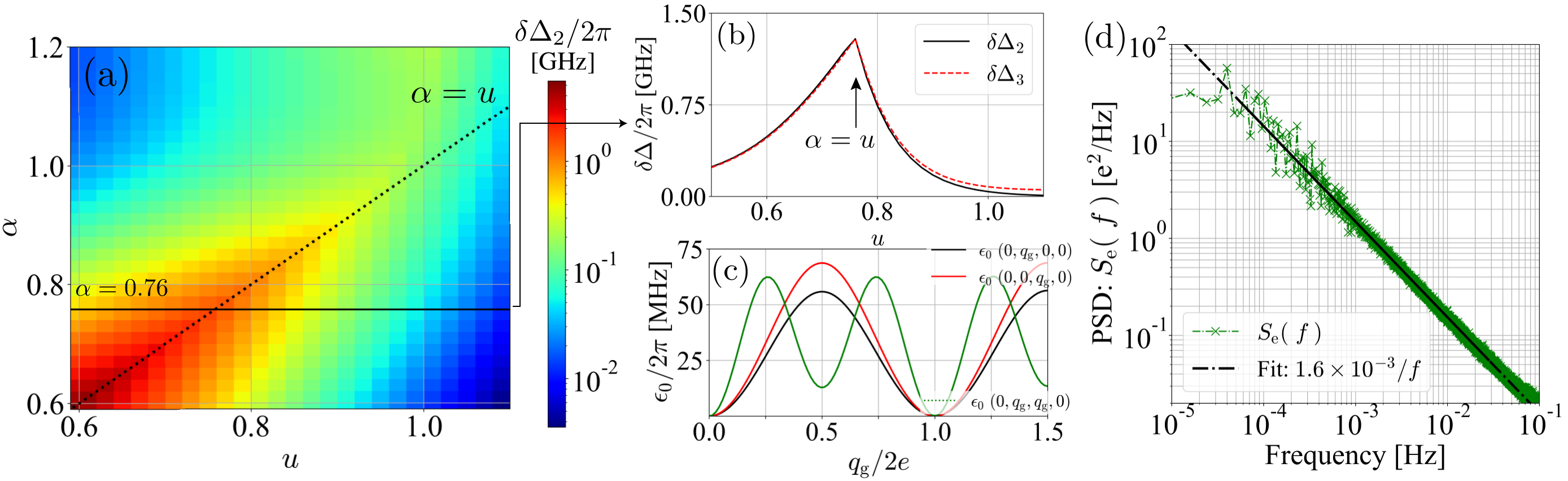}
\caption{
\add{(a) The split width of qubit energy gap frequency between two charge parity states, $\hbar\delta \Delta_2\equiv \hbar\Delta(0,e,0,0)-\hbar\Delta(0,0,0,0)$, plotted against the ratios of the $\alpha$- and $u$-junctions to the $v$-junction.
From the geometric symmetry of the circuit, $\delta \Delta_3\equiv \Delta(0,0,e,0)-\Delta(0,0,0,0)$ corresponds to the graph in (a) with the $u$-axis and $\alpha$-axis swapped.
(b) Cross section of (a) and $\delta\Delta_3$ at $\alpha=0.76$ plotted against $u$. When $\alpha<u<v=1$, we obtain $\delta \Delta_3>\delta \Delta_2$.
(c) Ground-state energy $\epsilon_0(q_\mathrm{g1},q_\mathrm{g2},q_\mathrm{g3},q_\mathrm{g4})$ of each charge parity state. We define $\epsilon_0(0,0,0,0)\equiv0$ as a reference point.}
(d) Power spectrum density of the charge offset fluctuation \add{obtained from}\erase{in} Fig.~\ref{Fig2}(c). 
The solid black line represents the $1/f$ fitting line and the dot\add{-dash} line is an extrapolation.
}
\label{Fig4}
\end{figure*}
\add{This result shows that the Aharonov--Casher effect appears as a change in $\Delta$ for the flux qubit: t}\erase{T}he Aharonov--Casher interference effect is most pronounced when the sizes of the two \add{smallest} junctions are equal and when the $\ejec$ ratio is small. 
The energy spectrum of the qubit is negligibly affected by \add{the} gate charge in islands 1 and 4 because the $\beta$-junction is much larger than \add{the} $u$- and $\alpha$-junctions. 
Additionally, the $\beta$-junction is shunted by the resonator's large inductance and capacitance, and the effective $\ejec$ ratio at the $\beta$-junction is \add{increased}\erase{suppressed} \add{in Eq.~\eqref{eq:Hq}}. Thereby, the \add{amount of change in}\erase{variation of} $\Delta$ with the charge in islands 1 and 4 is \add{suppressed and} lower than $10^{-9}~\si{\hertz}$, which is negligible in the measurement.
In contrast, the sizes of the $\alpha$-, \add{$u$-, and $v$-junctions are similar}\erase{$u$-junctions are close to that of the $v$-junction}, and these\erase{two} junctions play a major role\erase{s} in determining \add{$\hbar\Delta$ (}the energy gap of a qubit\add{)}~\cite{noauthor_see_nodate}.
Figure~\add{\ref{Fig4}(a) shows the dependence of how the qubit energy gap $\hbar\Delta$ varies due to quasiparticle poisoning in island 2 on the sizes of the $\alpha$- and $u$-junctions.
According to this result, in the flux qubit with several junctions, the split-width becomes larger when $\alpha$ and $u$ are similar, that is, it is a highly symmetric structure with respect to the $v$-junction.}
It is interesting to compare the present result with \add{those of} a previous work on \add{a} deep-strongly coupled qubit--resonator circuit~\cite{forn-diaz_ultrastrong_2017}, where \add{no}\erase{the} doubly split spectrum \add{was}\erase{were not} observed.
\add{Compared with the previous work~\cite{forn-diaz_ultrastrong_2017}, o}\erase{O}ur qubit has a lower $\ejec$ and \add{the size difference between the two smallest junctions is smaller}\erase{larger $\alpha$}.
The difference in the qubit design explains the difference \add{in}\erase{of} the observed spectrum.

\add{By} considering the dependence of $\Delta$ on the island charge in a flux qubit \add{[Fig.~\ref{Fig3}(c)]}, \add{one can see that} the observed spectrum splitting and its split-width fluctuation explain\erase{s} the presence of the two charge noises.
One of the noises originates from quasiparticles in islands tunneling back and forth through a junction, and this tunneling generates odd- and even-charge parity states.
If the system fluctuates between two parity states while it is probed with microwaves, the doubly split spectrum should be observed as a classical ensemble. 
\add{Although the quasiparticle tunneling rate at a Josephson junction depends on the density of quasiparticles, we do not expect it to be much longer than milliseconds on the basis of previous works}\erase{Quasiparticle tunneling events at a Josephson junction have a millisecond time scale}~\cite{riste_millisecond_2013,vool_non-poissonian_2014}, which is faster than the time for which microwaves are applied by the VNA (more than \add{20~ms}\erase{0.02s}) in Fig.~\add{\ref{Fig2}(c)}\erase{1(f)}.
The other noise is the environmental background electric field fluctuation around islands, which is not \add{caused by} a discretized charge such as a quasiparticle.
The charge \add{fluctuator}\erase{fluctuation} surrounding an island is ascribable to\erase{the large population of} electrons and holes \add{in the environment}, which generate an arbitrary offset gate charge through a capacitance $C_{\mathrm{g}i}$ as a fluctuation of the split width.
In other words, quasiparticle poisoning shifts the phase by $\pi$ in the cosine curve of \add{Figs.~\ref{Fig3}(b) and (c)}\erase{2(b)}, which corresponds to the shift between the solid and dash-dot lines, and the fluctuation of the background electric field on the sample can be understood as noise that continuously changes the split width.


\section{\add{Quasiparticle behavior}}\label{sec:QP}

The calculation results in \add{Figs.~\ref{Fig3}(b) and (c)}\erase{2(b)} indicate the appearance of four charge parity states, namely, $\mathcal{H}_\mathrm{q}(0,0,0,0)$,
$\mathcal{H}_\mathrm{q}(0,e,0,0)$,
$\mathcal{H}_\mathrm{q}(0,0,e,0)$, and
$\mathcal{H}_\mathrm{q}(0,e,e,0)$, which have different eigenenergies, in the spectrum measurement.
However, almost all traces of the iterated signals in Fig.~\add{\ref{Fig2}(c)}\erase{1(f)} show only one or two resonant modes, and the middle frequency of the split is constant. 
These observations indicate that the quasiparticle tunneling event mainly occurs in the $\alpha$- or $u$-junction, which is connected to the superconducting reservoir (the LC resonator in our circuit), and the state $\mathcal{H}_\mathrm{q}(0,e,e,0)$ is hardly generated.

\add{
Here, to infer at which junction quasiparticle tunneling occurs most frequently, we consider the energy difference of the system before and after one excess quasiparticle tunnels across a junction, which determines the tunneling probability of quasiparticles.
The energy change due to quasiparticle tunneling $\delta E_i$ is described using the initial and final charge state energies of the flux qubit ($\epsilon_i^\mathrm{initial}$ and $\epsilon_i^\mathrm{final}$) and the superconducting gaps of the initial and final islands of the quasiparticle position ($\Delta_\mathrm{sp}^\mathrm{initial}$ and $\Delta_\mathrm{sp}^\mathrm{final}$)~\cite{aumentado_nonequilibrium_2004}:
\begin{align}
    \delta E_i = \epsilon_i^\mathrm{final}-\epsilon_i^\mathrm{initial}+\Delta_\mathrm{sp}^\mathrm{final}-\Delta_\mathrm{sp}^\mathrm{initial}\,.
\end{align}
From the condition of double-angle shadow evaporation, the superconducting gap $\Delta_\mathrm{sp}$ in island 3 should be slightly larger than that in island 2 because the aluminum thickness of islands 1 and 3 is 30~nm and that of islands 2 and 4 is 40~nm, as shown in Figs.~\ref{Fig1}(d) and (e); thus, we expect the relation $\Delta_\mathrm{sp}^{(1)}=\Delta_\mathrm{sp}^{(3)}>\Delta_\mathrm{sp}^{(2)}=\Delta_\mathrm{sp}^{(4)}$~\cite{yamamoto_parity_2006}.
From the calculated ground-state energy of the flux qubit with each charge state in Fig.~\ref{Fig4}(c), when the size relation of the junctions is $v>u>\alpha$, we obtain $\epsilon_0(0,0,e,0)>\epsilon_0(0,e,0,0)>\epsilon_0(0,e,e,0)>\epsilon_0(0,0,0,0)\simeq\epsilon_0(e,0,0,0)\simeq\epsilon_0(0,0,0,e)$.}

\add{
Here, we consider four charge states $\vb{q}_\mathrm{g}^k=(e\delta_{k1},e\delta_{k2},e\delta_{k3},e\delta_{k4})$ and the transition between them, where $\delta_{kl}$ is the Kronecker delta and $k,l \in \qty{1,2,3,4}$ represent island indexes.
The ratio of the quasiparticle tunneling rate of island $k$ to that of $l$ is $\Gamma_{l\to k}/\Gamma_{k\to l}=\exp(-\delta E_i^{k\to l}/k_\mathrm{B}T)$ and when $\delta E_i^{k\rightleftarrows l}=0$, the quasiparticle tunneling rate is $\Gamma_{k\rightleftarrows l}^0\simeq E_\mathrm{c}/e^2R$, where $k_\mathrm{B}$ is the Boltzmann constant, $T$ is the environmental temperature, and $R$ is the room-temperature resistance of the junction~\cite{aumentado_nonequilibrium_2004,nakamura_quantitative_1996,tuominen_experimental_1992}.
Since $\Gamma_{k\rightleftarrows l}^0$ does not depend on the junction size, quasiparticle tunneling occurs more frequently as $\delta E_i^{k\to l}$ decreases.
Thus, the quasiparticle in the reservoir (LC resonator) tunnels to island 2 through the $u$-junction more frequently than to island 3 through the $\alpha$-junction, because the energy change of a quasiparticle tunneling from island 4 to 3 is much larger. 
When the excess quasiparticle is in island 2, it will tunnel to island 1 (reservoir) with high probability because the energy difference $\delta E_i$ for tunneling from island 2 to 3 is larger than that for tunneling from island 2 to 1.
Consequently, the quasiparticle tunneling mainly occurs at the $u$-junction~\cite{noauthor_see_nodate}.}

This \add{quasiparticle behavior}\erase{assumption} can also explain the fact that two states were observed in Refs.~\onlinecite{bertet_dephasing_2005,stern_flux_2014,bal_dynamics_2015}.
\add{This situation, in which the quasiparticles tunnel through one particular junction in the flux qubit, can also be applied to suppress decoherence due to quasiparticle poisoning by pumping quasiparticles away from the islands of a qubit using a series of $\pi$ pulses as reported in Ref.~\onlinecite{gustavsson_suppressing_2016}.}

\section{\add{$1/f$ background charge noise}}\label{sec:background}

Next, we focus on the background electric field fluctuation in detail. 
Figure~\add{\ref{Fig2}(d)}\erase{2(c)} shows the distribution of split values obtained by extracting the upper- and lower-branch frequencies from Fig.\add{~\ref{Fig2}(c)}\erase{1(b)}.
The \add{high}\erase{outstanding} count probability at 18$~\si{\mega\hertz}$ in Fig.~\add{\ref{Fig2}(d)}\erase{2(c)} corresponds to the charge offset around 0 (or mod $2e$)~\cite{noauthor_see_nodate}. 
Here, we apply two assumptions to obtain values of the island charge offset.
First, \add{as discussed in section~\ref{sec:QP},} the quasiparticle tunneling event mainly occurs at the \add{$u$-junction during the measurement of Fig.~\ref{Fig2}(c)}\erase{$\alpha$-junction in Fig.1(f) because the middle frequency of the split should be changed when tunneling occurs at two or more junctions}.
\erase{These quasiparticle picture can be applied to suppress decoherence due to quasiparticle poisoning by  pumping quasiparticles away from the islands of a qubit using a series of $\pi$ pulse as reported in Ref.~\onlinecite{gustavsson_suppressing_2016}.}
Second, \add{the background charge fluctuation also mainly affects the same island (island 2), because if the charge fluctuation in island 3 is greater than or equal to that in island 2}\erase{the charge fluctuation in the island 3 is much larger than that in the island 2 considering the area difference of the islands. If the charge fluctuation in islands 2 and 3 are comparable and has no obvious relation}, \add{the change in the} middle frequency \add{will}\erase{change would} be larger\add{~\cite{comment}}.
Then, charge offset values are obtained by converting $\delta \omega_{20}$ to the island charge \add{$q_\mathrm{g2}=0$ -- $0.5e$}\erase{$q_\mathrm{g3}=0$ -- $0.25e$} using the cosine curve dependence as shown in Fig.~\add{\ref{Fig3}(b)}\erase{2(b)}.
Since the periodicity is $2e$, any value above \add{$q=0.5e$}\erase{$0.25e$} falls in the range [$0$, \add{$0.5e$}\erase{$0.25e$}] (aliasing).   

The calculated power spectrum density (PSD) $S_q\qty(\,f\,)$ for this charge offset is shown in Fig.~\add{\ref{Fig4}(d)}\erase{Fig.2(d)}, which shows $1/f$ dependence with $S_q\qty(1~\si{\hertz})=(\add{4.06}\erase{1.99} \times10^{-2}~e/\sqrt{\si{\hertz}})^2$ from the intercept of the fitting function.
The value of $S_q\qty(1~\si{\hertz})$ is much larger than a typical value for a single-electron transistor (SET)\add{~\cite{schoelkopf_radio-frequency_1998,verbrugh_optimization_1995}}\erase{\cite{schoelkopf_radio-frequency_1998}} and a CPB~\cite{astafiev_quantum_2004}\add{, which have the $1/f$ noise characteristic}, but \add{it is} close to the value for transmon\add{s, which have a $1/f^{1.7\mathrm{-}2.0}$ noise characteristic~\cite{riste_millisecond_2013,serniak_hot_2018,christensen_anomalous_2019}}\erase{\cite{christensen_anomalous_2019}}.
The relatively large background charge value might be due to the island size, sample quality, and materials.
\erase{For example, in Ref.~\onlinecite{verbrugh_optimization_1995}, it was reported that the charge noise increases with the island size in SET devices.}

\section{\add{Conclusion}}

\erase{In conclusion,} We have observed a peculiar doubly split spectrum in ultrastrongly coupled qubit--resonator systems, which is caused by the charge fluctuation of superconducting islands. 
Fitting with the Rabi Hamiltonian revealed that the parameter fluctuation and the doubly split spectrum originate from neither the TLS nor the environmental parasitic modes. 
The analysis of the circuit Hamiltonian and the dependence of its energy on the island charge explain that the splitting of the energy spectrum originates from the fluctuation of the number of quasiparticles on islands. 
Moreover, the fluctuation of the split width is caused by the fluctuation of the background electric field. 
The design with a low $\ejec$ and the \add{two small junctions having similar sizes are the}\erase{a large $\alpha$ is the} reason\add{s why}\erase{that} our flux qubits \add{are susceptible to}\erase{have exhibited the fluctuation from the} charge noise.
A noteworthy point is that the middle frequency of the split is stable for a few days, \add{indicating that quasiparticles mainly poison one particular island and that the background electric fluctuation also mainly affects the same island}\erase{making it possible to convert the frequency split to a charge offset}.
\add{This island (island 2) was inferred from the energy difference between charge states.
When the middle frequency does not change, the frequency split can be converted to the charge offset.}
The PSD of the background electric field fluctuation shows $1/f$ dependence and a larger value of $S_q\qty(1~\si{\hertz})$ than the conventional SET and CPB.
We also showed \add{that} the monitoring of the island gate charge \add{of an ultrastrongly coupled system} by single-tone spectroscopy\erase{of an ultrastrongly coupled system, which} is helpful for evaluating the behavior of quasiparticles.

\section*{Acknowledgement}
We thank S. Shirai, S. Watabe, R. Wang, S. Kwon, Y. Zhou, T. Miyanaga, and T. Yoshioka for their thoughtful comments on this research. We also thank K. Kusuyama, K. Nittoh, and L. Szikszai for their supports of sample fabrication. This paper was based on results obtained from a project, JPNP16007, commissioned by
the New Energy and Industrial Technology Development Organization (NEDO), Japan. Supporting from JST CREST (Grant No. JPMJCR1676 and JPMJCR1775) and Moonshot R \& D (Grant No. JPMJMS2067) are also appreciated.

\bibliography{chargeOffsetUS} 

\begin{thebibliography}{53}
\bibitem [{com()}]{comment}%
  \BibitemOpen
  \href@noop {} {}\bibinfo {note} {\addsec{Based on the following three facts,
  we can conclude that the environmental electric field fluctuations are not
  uniform on each island and mostly affect island 2. First, quasiparticle
  tunneling rate does not depend on island size. Second, from the discussion of
  quasiparticle behavior in Section IV, the quasiparticles go back and forth
  between island 1 and island 2. Third, as an experimental fact, the middle
  frequency of the split does not change much in all four qubits measured.\\ As
  for the reason of the strong background electric fluctuation in island 2, we
  cannot exactly nail down. However, we can point out some possibilities, for
  example, only island 2 is accompanied by a small floating island right next
  to it [Fig. 1(d)], and the charge fluctuation there might have enhanced the
  charge fluctuation in island 2. Also, the much smaller island size of island
  2 compared to other islands might have enhanced the fluctuation.
  }\erasesec{We cannot conclude the reason of the strong background electric
  fluctuation in island 2. The background field fluctuation originated from the
  floating island next to island 2 or the micro-scale fluctuation that is
  averaged out to sufficiently small in the area of island 3 could mainly
  affect island 2.}}\BibitemShut {Stop}%
\end{thebibliography}

\ifarXiv
    

\section*{Supplementary material}

# Supplemental Material for "Quasiparticle tunneling and $1/f$ charge noise in ultrastrongly coupled superconducting qubit-resonator"

A. Tomonaga,[1,2] H. Mukai,[1,2] F. Yoshihara,[3] and J. S. Tsai[1,2]

[1]*Department of Physics, Tokyo University of Science,*
*1–3 Kagurazaka, Shinjuku, Tokyo 162–0825, Japan*

[2]*RIKEN Center for Quantum Computing (RQC), 2–1 Hirosawa, Wako, Saitama 351–0198, Japan*

[3]*Advanced ICT Research Institute, National Institute of Information and Communications Technology,*
*4-2-1, Nukuikitamachi, Koganei, Tokyo 184-8795, Japan*

## S1. Spectrum and Rabi model

We summarize the fitted data of each spectrum for four different qubits in two samples with the same design in Table SI, where all qubit-resonator systems are in the ultrastrong coupling regime.

In all figures showing spectra, the horizontal axis is converted to $\varepsilon/2\pi$ in the fitted Rabi (or circuit) Hamiltonian. The input power at the port of the sample holder is -130 dBm and the base temperature is below 10 mK. The two samples we report in this paper have the same design and their junctions are simultaneously fabricated on the same wafer by double-angle shadow evaporation of aluminum. The LC resonator is made from an interdigital capacitor and a 50 nm niobium thin-film inductance, which is fabricated on a high-resistance undoped silicon wafer.

Figure S1 shows another example of a spectrum and two sets of parameters in Rabi fitting. The spectrum is obtained by measuring the transmission signal compared with the input power from the VNA ($S_{21}$ measurement). The split width highly fluctuates with time (shown in main text), but its time scale is dozens of minutes or longer; thus, the spectrum was measured when the split width was stable. In Fig. S1, there are 4.14 and 4.59 GHz resonant modes, both of which are weakly coupled to the system. They may be parasitic standing waves originating from sample ground planes or the measurement environment including the sample holder, connectors, and

TABLE SI. Fitted parameters of each qubit in Rabi model. Split $\delta\Delta$ is equal to the difference between the fitted qubit gaps for two branches: $\Delta_{\text{blue}} - \Delta_{\text{green}}$. The upper branch corresponds to, for instance, the blue part in Fig. S1.

| Sample label | Split [MHz] $\delta\Delta/2\pi$ | Fitted parameters [GHz] $\Delta_{\text{blue}}/2\pi$ | $g/2\pi$ | $\omega_{\text{r}}/2\pi$ |
|---|---|---|---|---|
| A.Q1 | 70 | 1.138 | 2.226 | 4.429 |
| A.Q2 | 66 | 0.863 | 2.225 | 4.462 |
| B.Q1 | 122 | 1.338 | 2.873 | 4.281 |
| B.Q2 | 153 | 1.079 | 1.922 | 4.333 |

other microwave components, but these parasitic modes are not a source of doubly split spectrum as mentioned in main text and do not affect the analysis of charge fluctuations.

In the spectra shown in Fig. S1, a large doubly split is observed at the $\omega_{20}$ transition, which correspond to the one-photon excitation of the resonator from the ground state of the system in the weaker coupling regime. Figure S2 shows the dependence of $\omega_{20}/2\pi$ on the coupling constant in Rabi Hamiltonian when two $\Delta$ have a hundred MHz difference. The frequency difference of red and black lines correspond to the split width of $\omega_{20}/2\pi$. When the coupling strength is small, the resonator is not signif-

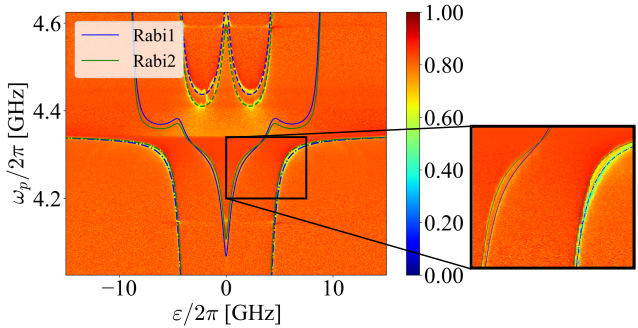

FIG. S1. Observed single tone spectrum of sample B.Q2 and fitted curves with the two different sets of parameters in TABLE SI. $\omega_{10}$ (dash-dot), $\omega_{20}$ (dash), and $\omega_{30}$ (solid) represent state transition frequencies in the Rabi Hamiltonian.

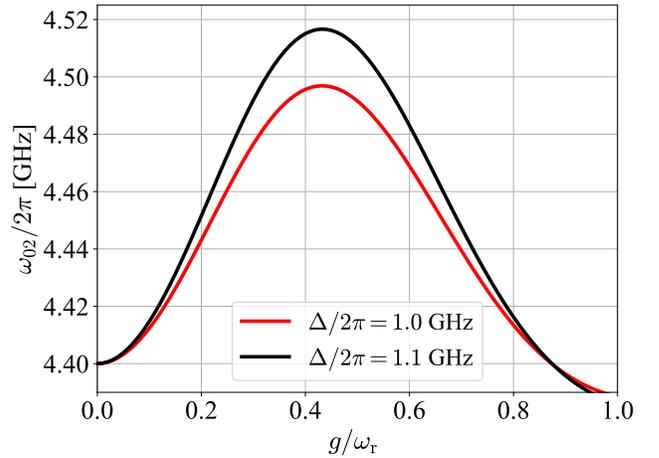

FIG. S2. The dependence of qubit gap $\Delta$ on the ratio of the coupling constant to the resonator frequency in the Rabi Hamiltonian. Parameters $\omega_{\text{r}}/2\pi = 4.4$ GHz, and $\varepsilon/2\pi = 1.64$ GHz are used.

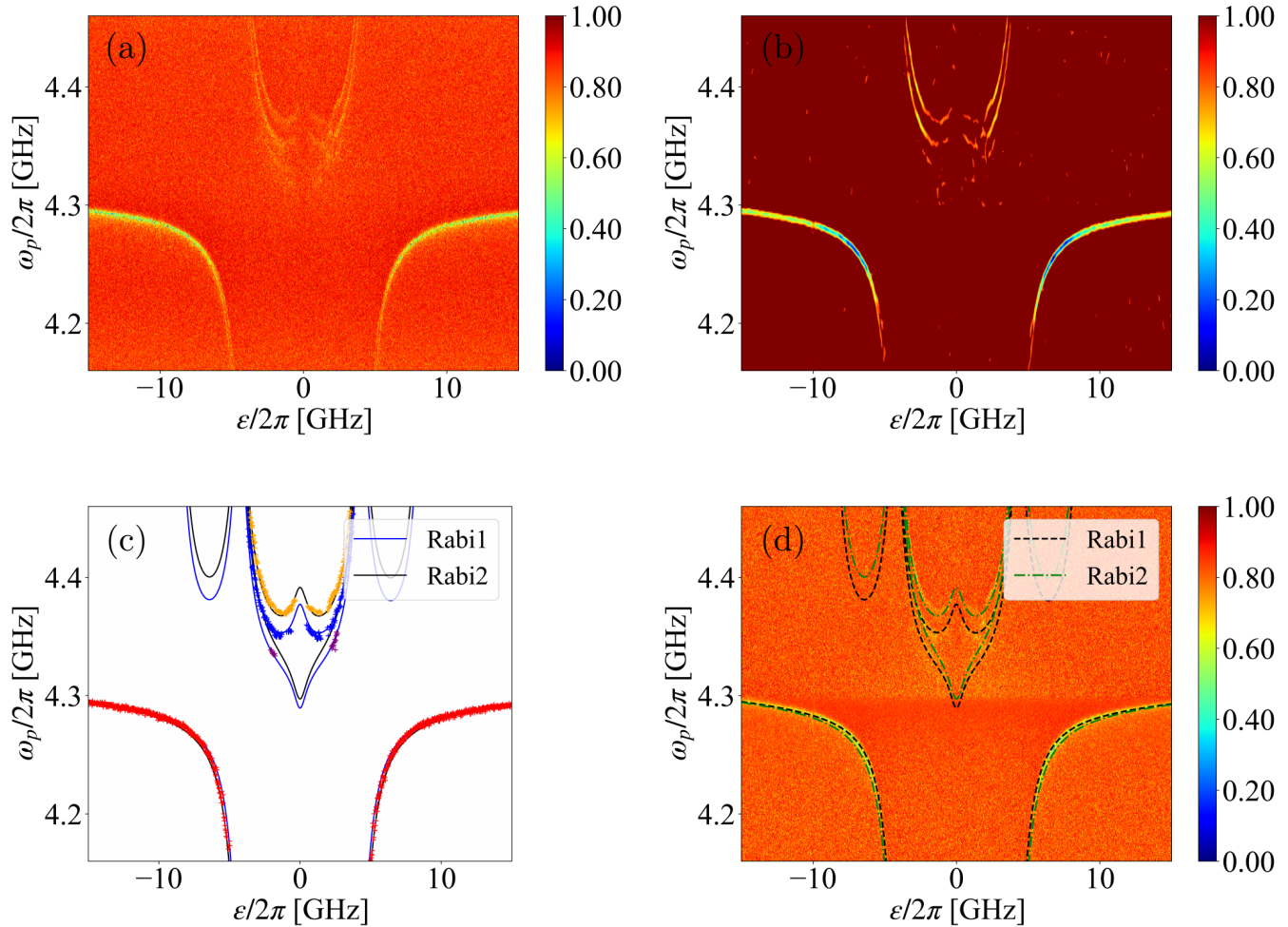

FIG. S3. (a) Original single tone spectroscopy data of sample B.Q1. (b) Filtered image of the data in (a). The "sato" function in "scikit-image" is used to emphasize the ridgelike structure. (c) Traces of each peak point in each branch of energy absorption by system state transitions. Each branches are not overlapped. (d) Fitted curves obtained by least-squares fitting of all branches using two sets of parameters in Rabi Hamiltonian, which shown in the TABLE SI.

icantly affected by the fluctuation of the qubit parameter $\Delta$, i.e., the splitting of $\omega_{20}/2\pi$ at $g/2\pi = 100$ MHz and $\varepsilon = 0$ is 200 kHz. On the other hand, in the ultrastrong coupling regime, the splitting in sample B.Q2 at $\varepsilon = 0$ is 49 MHz; thereby, the fluctuation of the ground state energy of the qubit has a significant effect on the transition frequency $\omega_{20}/2\pi$. The number of virtual photons of the resonator in the ground state of this system (sample B.Q2) is around 0.14, indicating that a qubit and a resonator are inseparable even in the ground state. In the Rabi model, when $4g^2/(\omega_r\Delta) > 1$, the eigenstate of the system shows a strong entanglement between the qubit and the resonator[1], and $4g^2/(\omega_r\Delta) = 3.16$ in sample B.Q2. The effective qubit gap $\tilde{\Delta}$ is reduced by the factor $\exp\left[-2(g/\omega_r)^2\right]$; thus, the split of $\omega_{20}$ will not be seen[1]. Our devices have the ratio $g/\omega_r$ around 0.5, which make high visibility of $\omega_{20}$ split.

## S2. Peak tracing

To fit the spectrum, the information of the frequency in each branch is required. From the noisy spectrum, we obtain peak points using various filtering methods. Figure S3 shows one method used to obtain peak points from the spectrum. Using the "scikit-image" Python package for image processing, we apply the ridge filter to the spectrum in Figure S3(a) and obtain the peak data area inside the contour lines with a determined threshold value[2,3]. Then, we take the peak data as the minimum point with each current bias in the filtered area [Fig. S3(b)].We also use the two-dimensional (time and frequency spaces) moving average and normalization as required for each spectrum before applying the ridge filter.

## S3. Circuit analysis

Here, we describe the circuit Hamiltonian calculation in detail. The branch flux across the circuit elements, which are junctions, inductance $L_r$, and capacitance $C_r$, follow Kirchhoff's voltage laws:

$$\varphi_\beta + \varphi_\alpha + \varphi_u + \varphi_v = \varphi_{\text{ext}}, \tag{S1}$$

$$\varphi_{cr} + \varphi_{\ell r} = \varphi_\beta, \tag{S2}$$

with the node fluxes defined by

$$\begin{aligned}
\varphi_{cr} &\equiv \varphi_5 - \varphi_4, \\
\varphi_{\ell r} &\equiv \varphi_1 - \varphi_5, \\
\varphi_\beta &\equiv \varphi_1 - \varphi_4, \\
\varphi_u &\equiv \varphi_2 - \varphi_1, \\
\varphi_v &\equiv \varphi_3 - \varphi_2, \\
\varphi_\alpha &\equiv \varphi_4 - \varphi_3 + \varphi_{\text{ext}}.
\end{aligned} \tag{S3}$$

The total Lagrangian of the circuit is described as

$$\mathcal{L}_{\text{tot}} = \mathcal{K}_J + \mathcal{K}_{cg} - \mathcal{U}_{Lr} - \mathcal{U}_J + \mathcal{L}_r, \tag{S4}$$

where

$$\mathcal{K}_J = \frac{C_J}{2}\left[\beta\dot{\varphi_1}^2 + u(\dot{\varphi_2} - \dot{\varphi_1})^2 + (\dot{\varphi_3} - \dot{\varphi_2})^2 + \alpha\dot{\varphi_3}^2\right], \tag{S5}$$

$$\begin{aligned}
\mathcal{K}_{cg} = \frac{C_{g1}}{2}\left(\dot{\varphi_1} - V_{g1}\right)^2 &+ \frac{C_{g2}}{2}\left(\dot{\varphi_2} - V_{g2}\right)^2 \\
&+ \frac{C_{g3}}{2}\left(\dot{\varphi_3} - V_{g3}\right)^2, \tag{S6}
\end{aligned}$$

$$\begin{aligned}
\mathcal{U}_J = -E_J[\beta\cos(\varphi_1) &+ u\cos(\varphi_2 - \varphi_1) \\
&+ \cos(\varphi_3 - \varphi_2) + \alpha\cos(\varphi_{\text{ext}} - \varphi_3)], \tag{S7}
\end{aligned}$$

$$\mathcal{L}_r = \frac{C_r}{2}\dot{\varphi}_{cr}^2 - \frac{1}{2L_r}\varphi_{cr}^2 + \frac{1}{2L_r}\varphi_1\varphi_{cr}, \tag{S8}$$

and

$$\mathcal{U}_{Lr} = E_{Lr}\varphi_1^2. \tag{S9}$$

Ignoring the constant part in $\mathcal{K}_{cg}$ [Eq. (S6)], the qubit kinetic energy part of the Lagrangian in Eq. (S4) becomes

$$\begin{aligned}
\mathcal{K}_q &\equiv \mathcal{K}_J + \mathcal{K}_{cg} \\
&= \frac{1}{2}\dot{\boldsymbol{\phi}}^T \mathbf{M}\dot{\boldsymbol{\phi}} - \dot{\boldsymbol{\phi}}^T \mathbf{q}_g, \tag{S10}
\end{aligned}$$

where $\boldsymbol{\phi} \equiv (\varphi_1 \ \varphi_2 \ \varphi_3)^T$ and the mass matrix is given as

$$\mathbf{M} = C_J\begin{pmatrix} \beta + u + \eta_1 & -u & 0 \\ -u & 1 + u + \eta_2 & -1 \\ 0 & -1 & \alpha + 1 + \eta_3 \end{pmatrix}. \tag{S11}$$

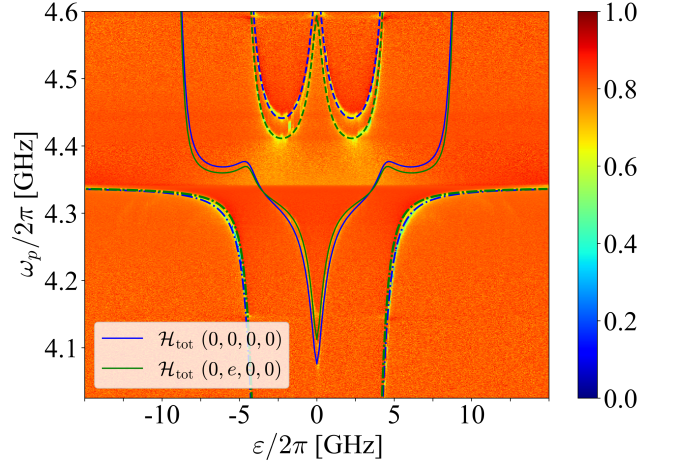

FIG. S4. Measured spectrum of sample B.Q2 and fitted curve using circuit Hamiltonian $\mathcal{H}_{\text{tot}}$. $\mathcal{H}_{\text{tot}}$ is fitted to the each transition frequency in the Rabi Hamiltonian of Fig. S1 using two different charge parity states. The fitted parameters are $E_J/h = 127$ GHz, $E_c/h = 3.72$ GHz, $L_r = 7.14$ nH, $\omega_r = 4.51$ GHz, $\alpha = 0.74$, $\beta = 2.24$, $u = 0.87$, and $\eta_{1,2,3} = 0.05$. The estimated junction area size ratios obtained from the SEM image (not shown) are $\alpha = 0.65$, $\beta = 1.92$, and $u = 0.89$.

Using the canonical conjugate $q_i = \partial L_{\text{tot}}/\partial\dot{\varphi_i}$ for $\dot{\varphi_i}$ and the charge basis $q_i' \equiv q_i + C_{gi}V_{gi}$, which represents the sum of Cooper pairs and the excess gate charge in an island, Eq. (S10) can be written as

$$\mathcal{K}_q = \frac{1}{2}\mathbf{q}'^T\mathbf{M}^{-1}\mathbf{q}' - \mathbf{q}'^T\mathbf{M}^{-1}\mathbf{q}_g, \tag{S12}$$

using $\mathbf{q}' = \mathbf{q} + \mathbf{q}_g = \mathbf{M}\dot{\boldsymbol{\phi}}$. Then, we obtain the total Hamiltonian of the circuit as

$$\begin{aligned}
\mathcal{H}_{\text{tot}} &= \mathbf{q}^T\dot{\boldsymbol{\phi}} - \mathcal{L}_{\text{tot}} \\
&= \frac{1}{2}\mathbf{q}'^T\mathbf{M}^{-1}\mathbf{q}' + \mathcal{U}_{Lr} + \mathcal{U}_J + \mathcal{H}_r \\
&= 4E_c\tilde{\mathbf{q}}'^T\tilde{\mathbf{M}}^{-1}\tilde{\mathbf{q}}' + \mathcal{U}_{Lr} + \mathcal{U}_J + \mathcal{H}_r, \tag{S13}
\end{aligned}$$

where $2e\tilde{q} = q$ and $C_J\tilde{\mathbf{M}} = \mathbf{M}$. The gate charge of each island is described as an offset on this basis.

To reduce the size of the matrix in numerical diagonalization, we deal with the qubit and resonator terms of the Hamiltonian in Eq. (S13) separately. First, we diagonalize the qubit Hamiltonian

$$\mathcal{H}_q \equiv 4E_c\tilde{\mathbf{q}}'^T\tilde{\mathbf{M}}^{-1}\tilde{\mathbf{q}}' + \mathcal{U}_{Lr} + \mathcal{U}_J, \tag{S14}$$

to obtain the eigenvectors. The Hamiltonian of the entire system [Eq. (S13)] is then expanded by the eigenvector $|i\rangle$ of $\mathcal{H}_q$, and we obtain

$$\begin{aligned}
\mathcal{H}_{\text{tot}} = \sum_i \hbar\omega_i|i\rangle\langle i| &+ \hbar\omega_r\left(a^\dagger a + \frac{1}{2}\right) \\
&- E_L\sum_{i,j}\langle i|\varphi_\beta|j\rangle|i\rangle\langle j|\left(a^\dagger + a\right), \tag{S15}
\end{aligned}$$

where $E_L = \hbar I_{zpf}\Phi_0$. To compute $\mathcal{H}_q$, we use the finite charge spaces of each node at least $(n_1, n_2, n_3) = (8, 5, 5)$. The sizes of these calculation spaces are determined as the values convergent within a 1 MHz difference, which is accurate enough to compute the spectrum. When we expand the total Hamiltonian in Eq. (S13) using the eigenvectors of $\mathcal{H}_q$, the size of the space in qubit $|i\rangle$ is up to $i = 10$ and the Fock state of the resonator requires up to 10 photons.

Figure S4 shows the spectrum of sample B.Q2 and the curve fitted using two charge parity states $\mathcal{H}_{tot}(0, 0, 0, 0)$ and $\mathcal{H}_{tot}(0, e, 0, 0)$. Although we cannot directly determine the amount of gate charge on each island, the fitting outcome reproduces the spectrum well and indicates that the charge offset of islands can make this splitting. We also confirmed that the circuit Hamiltonian reproduces the spectrum of the ultrastrong coupling system as well as the Rabi model. In the fitting parameters, the resonator inductance $L_r$ depends on the state and bias point of the next qubit (B.Q1 in Fig. S4) as discussed in the section S5.

## S4. Circuit and Rabi model Hamiltonians

The most obvious difference between the Rabi model and the circuit model is that the artificial atom is considered as two-level system or multi level system. To compare these two models, we use only two levels, $|0\rangle$ and $|1\rangle$, to calculate Eq. (S15), then, the two-level approximated circuit Hamiltonian reads

$$\mathcal{H}_{2L} = \hbar\begin{pmatrix} \omega_0 & 0 \\ 0 & \omega_1 \end{pmatrix} + \hbar\omega_r\left(a^\dagger a + \frac{1}{2}\right) + \hbar\begin{pmatrix} g_z & g_x e^{-i\xi} \\ g_x e^{i\xi} & -g_z \end{pmatrix}\left(a^\dagger + a\right),$$ (S16)

where $\{\omega_0, \omega_1, g_z, g_x, \xi\} \in \mathbb{R}$. The first term can be written as $\hbar\delta\omega\sigma_z/2$ by shifting the reference point, where $\hbar\delta\omega \equiv \omega_0 - \omega_1$. Off-diagonal components of the coupling term are mapped to real numbers using rotation about z-axis of $\xi$: $e^{-i\xi\sigma_z/2}$. Then, Eq. (S16) is written as

$$\mathcal{H}_{2L} = \frac{\hbar}{2}\delta\omega\sigma_z + \hbar\omega_r\left(a^\dagger a + 1/2\right) + \hbar\sqrt{g_z^2 + g_x^2}(\sigma_z\cos\theta_g + \sigma_x\sin\theta_g)\left(a^\dagger + a\right),$$ (S17)

where $\theta_g = \arctan g_x/g_z$, which is the same form as Refs. 4 and 5. We define the qubit energy gap $\Delta_{2L}$ as the minimum of $\omega_1 - \omega_0$ in Eq. (S17) and the rotation matrix as

$$U_\theta = \begin{pmatrix} \cos\theta/2 & -\sin\theta/2 \\ \sin\theta/2 & \cos\theta/2 \end{pmatrix},$$ (S18)

where $\theta = -\arctan\Delta_{2L}/\varepsilon$ and $\varepsilon \equiv \sqrt{\delta\omega^2 - \Delta^2}$. Assuming that $\theta \simeq \theta_g$ in the neighborhood of $\varphi_{ext}/2\pi = 0.5$,

and performing the unitary rotation using Eq. (S18), the Hamiltonian in Eq. (S17) is found to be

$$\mathcal{H}_{2L} \simeq \frac{\hbar}{2}(\varepsilon\sigma_z + \Delta_{2L}\sigma_x) + \hbar g_{2L}\sigma_z\left(a^\dagger + a\right) + \hbar\omega_r\left(a^\dagger a + \frac{1}{2}\right),$$ (S19)

where $g_{2L} = \sqrt{g_x^2 + g_z^2}$. The coefficients for each term are calculated as $g_{2L}/2\pi = 1.829$ GHz, $\Delta_{2L}/2\pi = 1.080$ GHz, and $\omega_r/2\pi = 4.584$ GHz from the circuit parameters used in Fig. S4. The form of the Hamiltonian in Eq. (S19) is the same as that of the Rabi model, but the resonator frequency in Eq. (S19) differs from the fitting parameters of the Rabi model ($\omega_r/2\pi = 4.333$ GHz in Fig. S1). Unlike the Rabi model, we arbitrarily choose the bare resonant frequency as the coefficient of $a^\dagger a$, consisting of an inductance and a capacitance, and the resonator frequency depends on the resonator basis.

From the numerical diagonalization of Eqs. (S15) and (S19), the neglected coupling terms between the qubit higher energy levels and the resonator in Eq. (S19) reduce the effective resonator frequency in the Rabi model. The $A^2$ term, which is prominent in ultrastrong coupling and widely known in atomic physics[6-8], has also the effect of reducing the resonator frequency by performing a unitary transformation[9].

## S5. Two-qubit interaction

In this study, the measured samples have two qubits connected to a single LC resonator, and the total Hamiltonian is described by

$$\mathcal{H}_{2Q} = \mathcal{H}_q^1 + \mathcal{H}_q^2 + \mathcal{H}_r^2.$$ (S20)

The form of $\mathcal{H}_q^{1,2}$ is the same as Eq. (S14), and

$$\mathcal{H}_r^2 = \frac{C_r}{2}\dot{\phi}_{cr}^2 + \frac{1}{2L_r}\phi_{cr}^2 - \frac{1}{2L_r}(\phi_{\beta1} + \phi_{\beta2})\phi_{cr},$$ (S21)

where $\phi_{\beta1}$ and $\phi_{\beta2}$ are node fluxes of the $\beta$-junctions of two qubits. Therefore, the two-qubit-coupled one-resonator system can be calculated in the same way as the single-qubit-coupled case [Eq. (S15)] by expanding the Hamiltonian of the entire system using the eigenvectors of the two qubits.

Here, we focus on the spectrum of qubit 1 ($\mathcal{H}_q^1$) and we bias qubit 2 ($\mathcal{H}_q^2$) away from the optimal point. Because the discussion in Supplementary material S4 can only be applied when the magnetic flux through the loop is close to half-integer multiples of a flux quantum, qubit 2 should be considered in the circuit model. However, the qubit ground to first excite state transition frequency ($\omega_{10}/2\pi$ in $\mathcal{H}_q$) away from half-integer multiples of a flux quantum is more than 30 GHz in our system and such a high transition energy suppresses the excited-state population. Thereby, qubit 2 is basically in the ground state

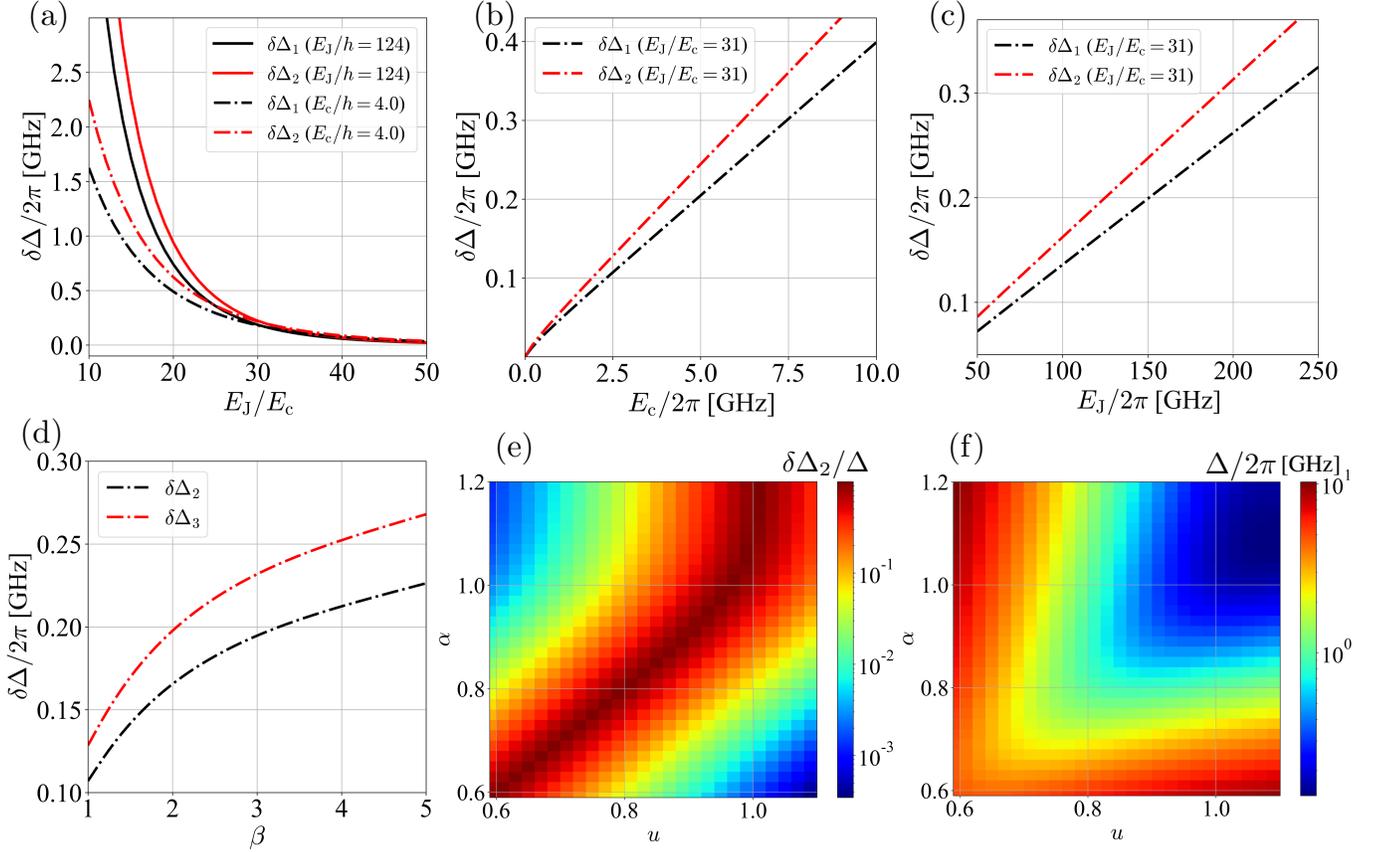

FIG. S5. The dependence of split width of qubit gap with circuit parameters $E_J/h = 124$ GHz, $E_c/h = 4.02$ GHz ($E_J/E_c = 30.7$), $\omega_r/2\pi = 4.68$ GHz, $L_r = 6.84$ nH, $\alpha = 0.76$, $\beta = 2.02$, $u = 0.90$, and $\eta_{1,2,3} = 0.12$. (a) The dependence on the ratio $E_J/E_c$. To change the $E_J/E_c$ ratio, we use the constant value of $E_J/h = 124$ GHz for solid line and $E_c/h = 4.02$ GHz for dash-dot line. (b)(c) $E_c$ and $E_J$ dependence with the constant $E_J/E_c$ ratio. (d) $\beta$-junction size dependence. (e) The splitting ratio to the qubit energy gap $\hbar\Delta$ depending on $\alpha$ and $u$ obtained from (f) and the Fig. 4(a) in the main text. (f) The qubit energy gap $\hbar\Delta$ depending on $\alpha$ and $u$.

and can be dealt with a classical inductance. Also from the numerical calculation of Eq. (S21), $\mathcal{H}_q^2(\varphi_{\text{ext}} \simeq 0)$ does not affect the spectrum shape of qubit 1. Consequently, in the present analysis, the effect of the qubit on the other side of the resonator can be treated as a change in resonator frequency, including the fitting parameters $L_r$ and $C_r$ to reproduce the spectrum in Fig. S4.

## S6. Charge sensitivity

Figure S5 shows the dependencies of the split width of qubit energy gap frequency between two charge parity state in islands 2 or 3 on each circuit parameters. The split width of qubit energy gap frequency is defined as $\delta\Delta_2 \equiv \Delta(0,0,0,0) - \Delta(0,e,0,0)$ and $\delta\Delta_3 \equiv \Delta(0,0,0,0) - \Delta(0,0,e,0)$, where $\Delta(q_{g1}, q_{g2}, q_{g3}, q_{g4})$ is the state transition frequency between the lowest two energy eigenstates in the qubit Hamiltonian $\mathcal{H}_q(q_{g1}, q_{g2}, q_{g3}, q_{g4})$ depending on islands gate charge. From Fig. S5(a), the $E_J/E_c$ ratio exponentially affect the split width and our

qubits have around 30 $E_J/E_c$ ratio, which is in the offset charge sensitive regime ($E_J/E_c < 50$)[10]. Also our qubits have closer $\alpha$ and $u$, these values makes higher visibility of doubly split in Fig. 4(a) in the main text. From results in Figs. S5(e) and (f), qubit energy gap is suppressed and splitting rate is pronounced when $\alpha$ and $u$ are similar.

## S7. Quasiparticle existence probability

We derive the probability that a quasiparticle exists on each island $P_k$ ($k \in \{1, 2, 3, 4\}$), which is calculated from $\Gamma_{l \to k}/\Gamma_{k \to l} = \exp(-\delta E_i^{k \to l}/k_B T)$ using the detailed balance at the ground state[11]. Here, we consider that only single non-equilibrium quasiparticles exist in the circuit and ignore their generation process. The exact value of the superconducting gap in our device cannot be determined, but in the aluminum thin film, $\Delta_{\text{sp}}(30 \text{ nm}) - \Delta_{\text{sp}}(40 \text{ nm})$ is around 0–10 GHz according to a previous work[12]. A schematic energy diagram of each charge state is drawn in Fig. S6(a) for the case of

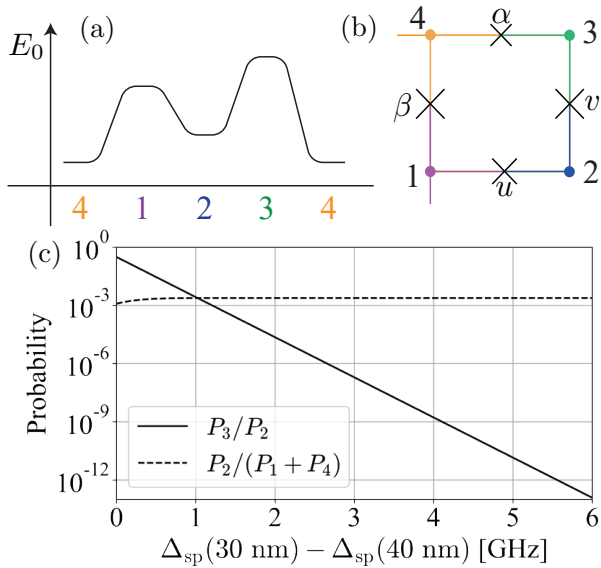

FIG. S6. (a) Schematic energy diagram of each charge state with one excess quasiparticle on each island. (b) Circuit diagram clarifying the location of each island. (c) Ratios of the probability that a quasiparticle exists on island 3 to that on island 2 and that a quasiparticle exists on island 2 to sum of that on islands 1 and 4 plotted against $\Delta_{\rm sp}(30~{\rm nm}) - \Delta_{\rm sp}(40~{\rm nm})$, where $\epsilon_0(0, e, 0, 0) = 0.56$ and $\epsilon_0(0, 0, e, 0) = 0.69$ GHz.

$\Delta_{\rm sp}(30~{\rm nm}) - \Delta_{\rm sp}(40~{\rm nm}) > \epsilon_0(0, e, 0, 0)$. The probability ratios of a quasiparticle existing on each island are shown in Fig. S6(c). When only a single quasiparticle exists in the circuit, $P_2/(P_1 + P_4)$ corresponds to the observation probability ratio of the lower branch to the upper branch in the spectrum measurements of Figs. 2(a) and (b). Although these probabilities should strongly depend on the number of quasiparticles in the actual circuit system, in the case of s single quasiparticle, the middle frequencies should be stable in the iteration measurement in Fig. 2(b) in the main text when $P_3/P_2$ is very small (for example, $\Delta_{\rm sp}(30~{\rm nm}) - \Delta_{\rm sp}(40~{\rm nm}) > 2$ GHz).

## S8. Charge distribution

Here, we summarize the method to obtain the power spectrum density of background charge noise in Fig. 4(d) in the main text. First, we use the two-dimensional (time and frequency spaces) moving average for Fig. 2(c) in the main text, in detail the signals for 4 intervals (12 s) and 5 frequency points (2.5 MHz) are averaged. Second, to obtain two resonant modes ($f_0^1$ and $f_0^2$), which are shown as small black dots in Fig. 2(c) in the main text, in iterated each trace, we fit each signal ($S_{21}$) using the

equation

$$S_{21} = S_{21}^0 + |S_{21}^1| + |S_{21}^2|, \qquad (S22)$$

where $S_{21}^0$ represents a offset signal level, and $S_{21}^k$ ($k \in 1, 2$) is written as[13,14]

$$S_{21}^k(f) = 1 - \frac{e^{i\phi^k} Q_{\rm L}^k/|Q_{\rm e}^k|}{1 + 2i Q_{\rm L}^k(f/f_0^k - 1)} \qquad (S23)$$

for each resonant mode. $Q_{\rm L}^k$, $Q_{\rm e}^k$, and $\phi^k$ represent loaded quality factor, external quality factor, and phase of $Q_{\rm e}^k$, which originates from the impedance mismatch between input and output, respectively. Figure 2(d) in the main text shows the split width $\delta f \equiv |f_0^1 - f_0^2|$ distribution, which is written as $\delta\omega_{20}/2\pi$ in the main text, obtained from this fitting. Fitted frequencies are affected by the electric signal fluctuation, so thus small charge fluctuations cannot be distinguished from the signal fluctuations. Also when the two modes are close, these two close modes are not easy to be distinguished and the obtained frequencies ($f_0^1$ and $f_0^2$) are largely affected by a fitting ambiguity and a signal noise. This is the reason why $\delta f = 0$ has high possibility to be seen in the distribution [Fig. 2(d) in the main text]. However large gate charge jumps (much larger frequency jump than a full width at half maximum of resonant modes) should be well distinguished from other noises such as a signal noise and a magnetic flux noise. Finally we obtain values of charge fluctuation $[0, 0.5e]$ from $\delta f$ [18 MHz, 0] using the cosine curve of $\omega_{20}/2\pi$ in Fig. 3(b) in the main text.

Thereby, we obtained PSD as the discrete Fourier transform of the autocorrelation function for measurement data. The autocorrelation function $R(j)$ of the discrete data $v(i)$ ($i$, $j \in \mathbb{N}$ represent measurement indexes) are defined by

$$R(j) = \frac{1}{N} \sum_{i=1}^{N} v(i)v(i+j), \qquad (S24)$$

where $N$ is the half amount number of the data and $j \in [1, N]$.

Figure S7(a) shows the power spectrum density (PSD) of frequency difference $f_0^1 - \bar{f}_0^1$, $f_0^2 - \bar{f}_0^2$, and $f_{\rm m} - \bar{f}_{\rm m}$, where $f_0^1 \gtreqqless f_0^2$, $f_{\rm m} \equiv (f_0^1 + f_0^2)/2$, and $\bar{f}$ represent the averaged frequency. The fluctuation of the middle frequency of the split originate from a signal noise and a magnetic flux noise. Figure S7(b) shows the PSD of the measured signal in iteration measurement [shown in Fig. S7(c)] at $f = 4.4875$ GHz close to $\omega_{20}/2\pi$, where there is no resonant mode near by. The signal PSD clearly shows the $1/f$ dependence, however, we cannot conclude whether the signal noise originates from a charge fluctuation on a sample or a charge fluctuation are raised by the $1/f$ signal noise originates from a measurement environment such as cables and amplifiers in this experiment.

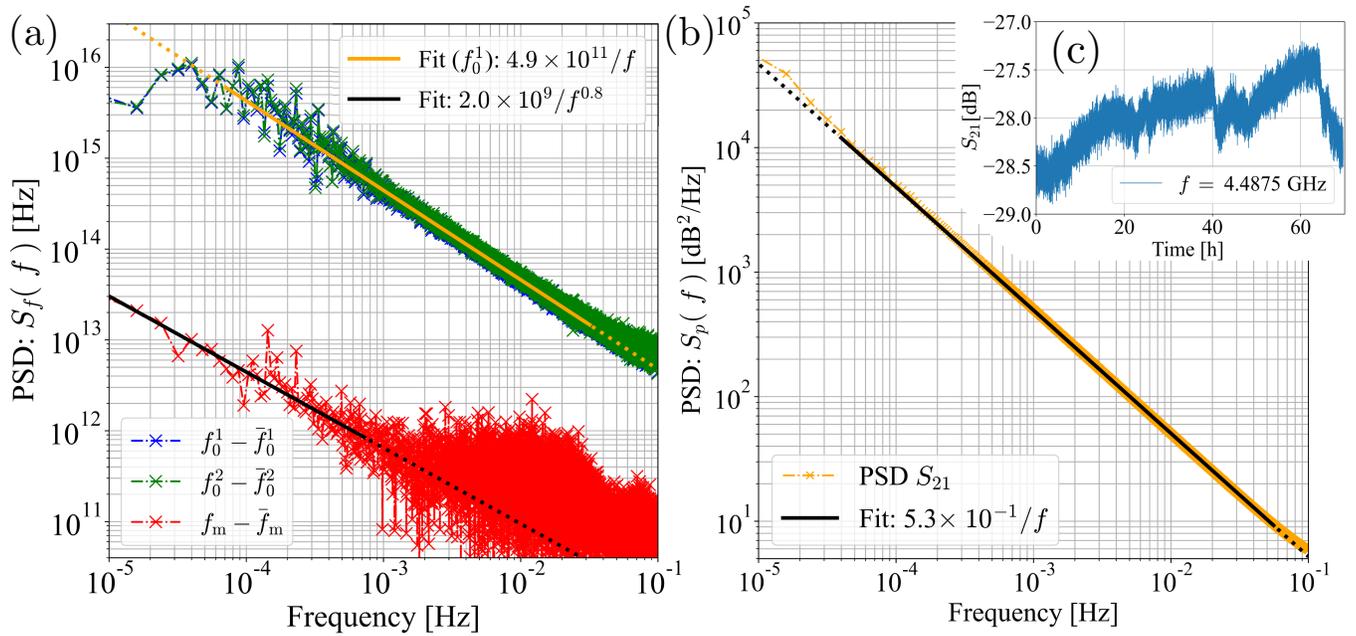

FIG. S7. (a) Power spectrum densities for the upper- and lower- branch frequencies fluctuation and the middle frequencies fluctuation of the split obtained from Fig. 2(c) in the main text. Solid lines represent liner fitting and dots represent the extrapolation area of the fitting. (b) Power spectrum density of the fluctuation of the scattering parameter $S_{21}$ at $f = 4.4875$ GHz from the measured signal shown in (c). (c) The signals obtained from iteration measurement [Fig.2(c) in the main text] at 4.4875 GHz.


\fi

\end{document}